\documentclass[aps,prx,onecolumn,notitlepage,superscriptaddress,11pt,tightenlines]{revtex4-2}
\usepackage[english]{babel}
\usepackage[T1]{fontenc}
\usepackage{graphicx}
\usepackage{amsthm,stmaryrd,amssymb} 
\usepackage{mathtools}

\usepackage{algorithm}
\usepackage{algpseudocode}\usepackage{enumerate}

\usepackage{tikz}
\usetikzlibrary{arrows.meta,calc,decorations.markings}
\usepackage{booktabs}

\usepackage{hyperref}

\DeclareMathOperator*{\ave}{\mathbb{E}}

\newcommand{\ket}[1]{\vert{#1}\rangle}

\newcommand{\bra}[1]{\langle{#1}\vert}

\newcommand{\bs}[1]{\boldsymbol{#1}}
\newcommand{\iu}{{i\mkern1mu}}

\DeclarePairedDelimiter\norm{\lVert}{\rVert}

\begin{document}

\title{Overshifted Parameter-Shift Rules: \\ Optimizing Complex Quantum Systems with Few~Measurements} 

\author{Leonardo Banchi} \email{leonardo.banchi@unifi.it}
\affiliation{Department of Physics and Astronomy, University of Florence,
via G. Sansone 1, I-50019 Sesto Fiorentino (FI), Italy}
\affiliation{INFN Sezione di Firenze, via G. Sansone 1, I-50019, Sesto Fiorentino (FI), Italy}

\author{Dominic Branford} 
\affiliation{Department of Physics and Astronomy, University of Florence,
via G. Sansone 1, I-50019 Sesto Fiorentino (FI), Italy}

\author{Chetan Waghela} 
\affiliation{Department of Physics and Astronomy, University of Florence,
via G. Sansone 1, I-50019 Sesto Fiorentino (FI), Italy}

\begin{abstract}
Gradient-based optimization is a key ingredient of variational quantum
algorithms, with
applications ranging from quantum machine
learning to quantum chemistry and simulation.
The parameter-shift rule provides a hardware-friendly method for evaluating
gradients of expectation values with respect to circuit
parameters, but its applicability is limited to circuits whose gate generators have
a particular spectral structure. 
In this work, we present a generalized framework 
that, with optimal minimum measurement overhead, 
extends parameter shift rules beyond this restrictive setting to
encompass basically arbitrary gate generator, possibly made of 
complex multi-qubit interactions with unknown spectrum and, in some settings,
even infinite dimensional systems such as those describing photonic devices 
or qubit-oscillator systems. 
Our generalization enables the use of more expressive
quantum circuits in variational quantum optimization and 
enlarges its scope by harnessing all the available hardware degrees of freedom. 
\end{abstract}

\maketitle

\section{Introduction} \label{sec:intro}
The rapid development of quantum computing hardware has
spurred a new era of algorithms designed to leverage the unique properties of
quantum mechanics. Among these, Variational Quantum Algorithms (VQAs) have
shown potential for applications in many areas, such as chemistry, materials
science, artificial intelligence and optimization~\cite{cerezo2021variational},
and represent one of the most promising approaches to harnessing near-term
quantum devices~\cite{bharti2022noisy}. VQAs operate by optimizing the
parameters of a quantum circuit to minimize a cost function that 
is estimated directly in hardware from measurable quantities. This
optimization process is often the most demanding part of the algorithm and 
typically relies on gradient-based methods~\cite{harrow2021low}.
Central to these methods is the estimation---directly in hardware---of gradients of expectation values with respect to circuit parameters, which then guide classical optimization routines. 

Among the techniques to directly estimate gradients in hardware,
parameter-shift rules have emerged as particularly elegant and practical methods~\cite{schuld2019evaluating,wierichs2022general,banchi2021measuring,wiersema2024here,pappalardo2025photonic,hoch2025variational,facelli2024exact}. 
They enable an unbiased estimation of the gradient with $\mathcal{O}(1)$ variance,
thus avoiding the large variance associated to finite difference methods, 
which typically result in convergence issues due to the excess stochastic noise. 
While remarkably effective for a wide class of quantum gates, the standard parameter-shift rule~\cite{schuld2019evaluating}
is often derived for and applied specifically to single-qubit gates. 
Further generalizations are still limited to special cases,
e.g.~gates generated by Hamiltonians with equally spaced frequencies~\cite{wierichs2022general}. 

These limitations restrict the design space of VQAs and may hinder their applicability to certain problems.
For instance, in quantum chemistry applications~\cite{bauer2020quantum}, 
variational circuit 
ans\"atze are typically built by exponentiating complex Hamiltonians, 
which often involve many-qubits and have a complex and possibly unknown
spectrum; here conventional parameter shift rules do not apply. 
Moreover, some novel or established quantum computing  architectures 
go beyond the qubit representation, 
e.g.~those based on qudits~\cite{roy2024qudit},
on the vibrational modes of ions~\cite{chen2023scalable}, 
on hybrid oscillator-qubit systems~\cite{crane2024hybrid} or 
on continuous variable systems, such as in 
photonic quantum computing 
\cite{larsen2019deterministic,giordani2023integrated,maring2024versatile,aghaee2025scaling}.
The Hamiltonians that describe these architectures may act on infinitely 
dimensional spaces where conventional methods do not apply. 
Although for some quantum algorithms there may exist some reasonable mapping---exact or approximate in some limit---from these systems to qubits,
for variational quantum algorithms this mapping is not necessary
and one can fully exploit the peculiarity of these more complex quantum systems. 

This paper addresses the limitations of known methods by presenting a
generalization of the parameter-shift rule,
applicable to basically \textit{any} arbitrary gate and quantum operation.
This generalization significantly expands the practical applicability of VQAs, enabling the use of more complex and expressive ans\"atze, as well as the use of all available degrees of freedom in different quantum computing architectures. 
Central to our analysis is the development of ``overshifted'' rules, where the 
number of parameter shifts is larger than what would be required by a mere 
counting argument. Within this extended space there are infinitely 
many solutions and, among them, we can select the parameter shift rule 
with minimum variance, which hence requires less measurement shots for 
estimating derivatives in hardware. The resulting problem for defining new parameter shift rules is convex and,
in some important limits, it can be approximated analytically and efficiently
even for large dimensional systems. 

We demonstrate that known parameter shift rules are special cases of our
generalized framework, and we provide the theoretical underpinnings for 
its optimality in terms of the total number of measurement shots.  
Our generalization not only broadens the theoretical foundations of variational quantum optimization but also provides a practical toolkit for implementing gradient-based learning in more expressive quantum models, opening pathways to improved algorithmic performance on near- and long-term quantum hardware.

The remainder of this paper is structured as follows: 
Sec.~\ref{sec:definition} provides the necessary background to formalize 
different parameter shift rules. 
Sec.~\ref{sec:overshifted} defines overshifted parameter shift 
rules and the convex optimization problem to find the ones with 
minimum variance, discussing also the connections with signal
processing. 
Sec.~\ref{sec:analytics} introduces analytic approximations 
that are sometimes based on infinite or continuously many shifts. 
Numerical simulations and different applications are 
considered  in Secs.~\ref{sec:numerics} and~\ref{sec:applic}. 
Conclusions and further research directions 
are drawn in Sec.~\ref{sec:Conclusions}.

\section{Problem definition}\label{sec:definition}

We focus on parametric quantum circuits expressed as a cascade of gates 
\begin{equation}
  \ket{\psi(\bs\theta)} =\hat W_L e^{i \theta_L \hat H_L}\cdots e^{i \theta_2
  \hat H_2}\hat W_1 e^{i \theta_1 \hat H_1}\ket{\psi_0},
  \label{eq:psi_theta}
\end{equation}
where $\bs\theta=(\theta_1,\dots,\theta_L)$, 
$\theta_j\in \mathbb{R}$ with $j=1,\dots,L$ define the tunable parameters, $L$ is 
the number of parameters, $\hat H_j$ are Hermitian operators (e.g. ``Hamiltonians''), and $\hat W_j$ are constant gates.
We are interested in derivatives of expectation values  
$f(\bs\theta)=\bra{\psi(\bs\theta)}\hat O\ket{\psi(\bs\theta)}$, for a certain observable $\hat O$. 
Without loss of generality, we can focus on a single parameter $\theta_k$ for a certain $k$ 
and study $\partial f(\bs\theta) / \partial \theta_k$, with all other parameters constant. 
Computing the full gradient is then trivial by repeating the same procedure for all possible $k$. 
Fixing $\theta\equiv\theta_k$, we may focus on 
\begin{equation}
  f(\theta) \equiv f(\bs\theta)\Bigg|_{\substack{\theta_k=\theta, \\\theta_{j\neq k}=\mathrm{const}}} = 
  \bra\psi e^{-\iu\hat H_k\theta} \hat M e^{\iu\hat H_k\theta}\ket\psi,
  \label{eq:ftheta def}
\end{equation}
where we have dropped the dependence on $k$ to simplify the notation and set 
$\ket{\psi}= \prod_{j=1}^{k-1} \hat U_j \ket{\psi_0}$,
$\hat{M}= W_k^{\dagger} \left(\prod_{j=k+1}^L \hat U_j\right)^\dagger \hat{O}
\left(\prod_{j=k+1}^L \hat U_j\right) W_k $,
where $\hat{U}_j = \hat{W}_je^{i \theta_j \hat{H}_j}$.
Without loss of generality, we may assume diagonal $\hat{H}_k$ with eigenvalues 
$E^{(k)}_j$, since we can always reabsorb the diagonalizing unitaries in the constant 
gates $\hat{W}_{k}$ and $\hat{W}_{k-1}$, and $\ket{\psi_0}$ for $k=1$. 
Therefore, we may express $f(\theta)$ as a Fourier-like series
\begin{equation}
  f(\theta) = \sum_{\omega \in \Omega} 
  M_\omega e^{\iu \omega \theta} 
  \label{eq:f_trig}
\end{equation}
where
\begin{equation}
  \Omega =\{ E^{(k)}_j - E^{(k)}_i \mathrm{~for~}i,j=1,\dots,N_k\},
  \label{eq:omega}
\end{equation}
is the set of ``beat'' frequencies (energy differences), $N_k$ is the number of distinct eigenvalues of $\hat H_k$ 
and the complex numbers  $M_\omega$ depend on the sum of operator elements 
$\bra{E^{(k)}_j} \hat{M} \ket{E^{(k)}_i}$ in the energy basis 
with $E^{(k)}_j - E^{(k)}_i =\omega $. 
Therefore, the total number of distinct frequencies satisfies $|\Omega| \leq \mathcal O(N_k^2)$. 
Since $f(\theta) \in \mathbb{R}$, the complex coeffients satify $M^*_{\omega}=M_{-\omega}$, 
and if $\omega\in\Omega$ then also $-\omega\in\Omega$. 

While \( \Omega \) is determined by the eigenvalues of \( \hat{H}_k \), as only eigenstates \( \{ \ket{E^{(k)}_j} \} \) with support on both \( \ket{\psi} \) and \( \hat{M} \) contribute to the \( M_{\omega} \) terms, a valid shift rule can be found for states \( \ket{\psi} \in \Psi \) and observables \( \hat{M} \in \mathcal{M} \) from only
\begin{equation}
	\Omega = \{ E^{(k)}_j - E^{(k)}_i | E^{(k)}_i, E^{(k)}_j \in \mathcal{E}_{\Psi}(\hat{H}_k) \cap \mathcal{E}_{\mathcal{M}}(\hat{H}_k) \},
  \label{eq:omega_minimal}
\end{equation}
where \( \mathcal{E}_{\Psi}(\hat{H}_k) = \{ E^{(k)}_j | \langle \psi | E^{(k)}_j \rangle \neq 0 \mathrm{~for~any~} \psi\in\Psi \} \) is the set of eigenstates which any \( \ket{\psi}\in\Psi \) have support on,
and \( \mathcal{E}_{\mathcal{M}}(\hat{H}_k) = \{ E^{(k)}_j | \hat{M} \ket{E^{(k)}_j} \neq 0 \mathrm{~for~any~} \hat{M}\in\mathcal{M} \} \) s the set of eigenstates which any \( \hat{M}\in\mathcal{M} \) have support on.

\subsection{Computing Gradients in the Quantum Hardware }\label{sec:gradients}

From the definition~\eqref{eq:f_trig}, it is now trivial to express the derivative as 
\begin{equation}
  f'(\theta) \equiv \frac{\partial f(\bs \theta)}{\partial \theta_k} = 
   \sum_{\omega \in \Omega} 
   M_\omega e^{\iu \omega \theta} \iu \omega.
   \label{eq:f_prime}
\end{equation}
However, for complex quantum circuits the coefficients $M_\omega$
may be very hard to compute with classical computers. Therefore, we try to express 
the derivative~\eqref{eq:f_prime} as a linear combination of evaluations of $f(\theta)$, 
since we already know how to estimate that quantity in a quantum computer: 
we sequentially apply the parametric gates 
and the constant gates $\hat{W}_j$ to first create the state~\eqref{eq:psi_theta},
and then measure the observable $\hat{O}$. 

In parameter shift rules~\cite{banchi2021measuring,wierichs2022general,pappalardo2025photonic,hoch2025variational}
we look for expressions like
\begin{equation}
	\frac{d f(\theta)}{d \theta} = \sum_{p} c_p f(\theta+\vartheta_p), 
  \label{eq:shift}
\end{equation}
where the real coefficients $c_p$ and shifts $\vartheta_p$ are unknown and must be obtained. 

Particular solutions to the above equation have been extensively studied in the literature~%
\cite{mitarai2018quantum,schuld2019evaluating,banchi2021measuring,wierichs2022general,wiersema2024here}.
The most popular one is for when $\hat{H}_k$ is a Pauli operator, with two distinct eigenvalues 
$E^{(k)}_j=\pm1$, for which two shifts $\vartheta_{\pm}=\pm\pi/4$, with weightings $c_{\pm}=\pm1$ are used.
For more general Hamiltonians, there are no explicit guidelines to define the shifts $\vartheta_p$ and the coefficients $c_p$.

By asking that~\eqref{eq:shift} and~\eqref{eq:f_prime} are equal for all possible 
values of $M_\omega$, which are unknown and hard to compute, we get the following 
expression 
\begin{align}
  \sum_{p} c_p e^{\iu \omega \vartheta_p} 
  &= \iu \omega, &
  \forall \omega &\in\Omega.
  \label{eq:linear_prob}
\end{align}
For general frequencies the above problem can be solved by 
the Nonequidistant Fast {Fourier} Transform, for which there are several numerical 
libraries, e.g.~\cite{knopp2023nfft}.

\subsection{Symmetric parameter shift rules}

We can simplify the above linear system, by considering only shifts that are symmetric around the origin, coming in pairs \( \pm \theta_p \) with equal and opposite coefficients.
In this case we require that 
\begin{equation}
  \frac{d f(\theta)}{d \theta} = \sum_{p=1}^P c_p [f(\theta+\vartheta_p) - f(\theta-\vartheta_p)],
  \label{eq:shift pos}
\end{equation}
where $P$ is the number of positive shifts $\vartheta_p>0$ and 
\begin{align}
  2 \sum_{p=1}^P c_p \sin(\omega \vartheta_p)
  &= \omega, &
	\forall \omega &\in\Omega^+,
  \label{eq:linear_prob pos}
\end{align}
where the real part of Eq.~\eqref{eq:linear_prob} is automatically satisfied for any symmetric parameter-shift rule of the form \eqref{eq:shift pos}.
The set $\Omega^+$ is the subset of $\Omega$ with strictly positive frequencies. 
Clearly $|\Omega| = 2|\Omega^+| +1$ since $\Omega$ contains all negative frequencies and the zero frequency. 
The linear system Eq.~\eqref{eq:linear_prob pos} may be solvable in general provided that the number of 
positive shifts $P$ satisfies 
\begin{align}
  P\geq & N & N:= |\Omega^+|.
  \label{eq:PNdef}
\end{align}
In general though there is no guideline to choose the shifts $\theta_p$. Since the functions $f(\theta\pm\vartheta_p)$ 
are typically estimated in a quantum hardware, one may naively guess that one should look for solutions 
with the smallest number of shifts, namely with $P=N$. However, since the optimal shifts $\theta_p$ are 
not known, we will show that it is in general beneficial  
to work in the overparametrized regime,
where the total number of shifts exceeds the number of constraints and the problems 
\eqref{eq:linear_prob},\eqref{eq:linear_prob pos}, have infinitely many solutions. 
We call the corresponding parameter shift rule ``overshifted'',
as more shifts than those required to solve Eq.~\eqref{eq:linear_prob} will be used.

As we will show, with the proper regularization, the solutions of many overshifted parameter shift rules will be 
sparse, namely the number of non-zero coefficients $c_p$ will still be small. Moreover, these solutions 
also minimize the variance and hence the number of 
measurements in the quantum hardware. 
In general, with $P\gg N$ we can find better parameter shift rules with several advantages. 

The ``infinitely overshifted'' limit $P\to\infty$ will also be quite useful: in that limit
parameter shift rules can be defined even without knowing the frequencies, 
but just by knowing an upper bound on the bandwidth 
\begin{equation}
  \Lambda = \max_{\omega\in \Omega} |\omega|.
  \label{eq:bandwidth}
\end{equation}
This is particularly useful when we are interested in gradients of complex quantum evolutions where 
the ``Hamiltonians'' $\hat H_\ell$ in Eq.~\eqref{eq:psi_theta} are complex, e.g.~many-body
Hamiltonians with unknown spectrum. Another interesting application is for dependent parameters, 
e.g.~for parameter sharing, as we define in the next section.

\subsection{Parameter sharing}\label{sec:sharing}

\begin{figure}[ht]
	\centering
    \begin{tikzpicture}
      \draw[very thick] (0,1) circle (1);
      \draw[domain=-2.5:-1.5,smooth,variable=\x,thick,red] plot ({\x},{0.8*exp(-((\x+2)^2)/0.05)});
      \draw[domain=-4:-3,smooth,variable=\x,thick,red] plot ({\x},{0.8*exp(-((\x+3.5)^2)/0.05)});
      \draw[domain=-5.5:-4.5,smooth,variable=\x,thick,red] plot ({\x},{0.8*exp(-((\x+5)^2)/0.05)});
      \draw[very thick] (-6,0) -- (4,0);
      \node[draw, rounded corners=2pt, inner sep=4pt, line width=0.8pt, fill=white] at (0,0) {$\theta$};
    \end{tikzpicture}
  \caption{Example circuit with hardware parameter sharing.
		Three time-bin encoded photonic qubits (red pulses) enter into a beam splitter, with angle $\theta$.
		The delay line length is tuned according to the time separation between each photons and applies the same beam splitting operation to each neighbouring pair. 
  }
  \label{fig:timebin}
\end{figure}
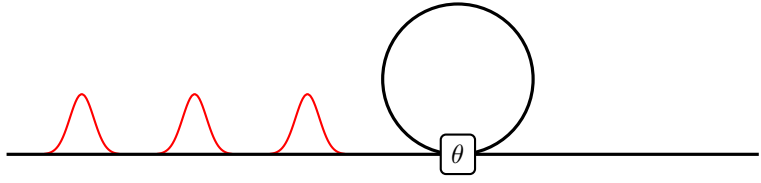

In some settings, some parameters in the quantum circuit definition~\eqref{eq:psi_theta} 
may be equal, either by design choices, e.g.~to approximate Floquet dynamics~\cite{ikeda2024robust}, 
or by hardware limitations. These limitations are quite common, for instance, in 
photonic quantum computing \cite{humphreys2013linear,larsen2019deterministic,madsen2022quantum}. 
An example is a beam splitter between time-bin 
encoded qubits, shown in Fig.~\ref{fig:timebin}, where the same parametric operation is 
applied to each neighbouring pair of qubits. 

The conventional approach would be to use the chain rule of derivatives,
but this breaks the time-translational invariance.  
We will show this with the following example 
\begin{equation}
  f(\theta_1,\theta_2) = \bra{\psi_0}e^{-\iu \theta_1 \hat H_1} \hat W e^{-\iu \theta_2\hat H_2} 
  \hat M e^{\iu\theta_2\hat H_2}\hat W e^{\iu\theta_1\hat H_1}\ket{\psi_0}. 
  \label{eq:f_theta_12}
\end{equation}
Suppose that hardware limitations force $\theta_1=\theta_2=\theta$, and assume $\hat{H}_1=\hat{H}_2$, 
as in the example of Fig.~\ref{fig:timebin}.
Then we can compute gradients 
using the chain rule and Eq.~\eqref{eq:shift} as 
\begin{equation}
  \frac{df}{d\theta} = \frac{\partial f}{\partial \theta_1}\frac{d\theta_1}{d\theta}
  + \frac{\partial f}{\partial \theta_2}\frac{d\theta_2}{d\theta} = 
  \sum_{p} c_p \left[f(\theta+\vartheta_p,\theta) 
  + f(\theta,\theta+\vartheta_p) \right].
  \label{eq:chain rule}
\end{equation}
However, to estimate gradients in hardware using the above formula we need to 
evaluate the circuit for non-symmetric parametrizations, which might not be possible. 
In the example of Fig.~\ref{fig:timebin}, this would require the dynamic modulation of $\theta$ 
at different times, which is more challenging than a static $\theta$. 

On the other hand, within the setting of Sec.~\ref{sec:definition}, we may 
express $f(\theta,\theta)$ using the expansion~\eqref{eq:f_trig} with 
\begin{equation}
  \Omega = \{ E_j + E_\ell - E_i -E_k \mathrm{~for~}i,j,k,\ell=1,\dots,N\},
\end{equation}
where $E_j$ are the eigenvalues of $\hat{H}_1$.
Thanks to Eq.~\eqref{eq:shift} we can now express all derivatives of $f(\theta,\theta)$ as a linear
combination of $f(\theta+\vartheta_p,\theta+\vartheta_p)$, which maintains the parametrization symmetry. 

More generally, suppose that that the parameter $\theta$ is shared among 
$T$ gates which are located at the layers $\ell_t \in [1,\dots,L]$. 
Then, we may use the solution of Sec.~\eqref{sec:gradients} with 
\begin{equation}
  \Omega =\left\{ \sum_{t,s=1}^T E^{(\ell_t)}_{j_t} - E^{(\ell_s)}_{i_t} ~\forall i_t,j_t=1,\dots,N\right\},
  \label{eq:omega_shared}
\end{equation}
The evaluation of the set $\Omega$ becomes prohibitively expensive for large $L$, as the 
number of frequencies increases exponentially with $L$. Nonetheless, we will show that some 
parameter shift rules only depend on the bandwidth Eq.~\eqref{eq:bandwidth}
which is much simpler to compute. For instance, in \eqref{eq:omega_shared} we can 
estimate the bandwidth $\Lambda$ by summing the largest frequencies in each layer.

\section{Overshifted Parameter-Shift Rule}\label{sec:overshifted}
Among the infinitely many solutions of~\eqref{eq:linear_prob} or \eqref{eq:linear_prob pos} in the 
overparametrized regime,
we need to select the most appropriate given the hardware constraints. 
The most severe one is typically the measurement overhead, namely the 
number of measurement shots to be performed in-hardware to have 
a reliable estimate of the gradient.

Motivated by this, we adapt an error estimation technique from Ref.~\cite{wierichs2022general}.
Suppose that the variance of $f(\theta)$ for any $\theta$ is bounded by $\sigma^2$.
Then if we estimate the $p$th element of the sum in~\eqref{eq:shift} with $S_p$ 
measurement shots, the variance of the gradient estimator is 
\begin{equation}
  \operatorname{Var}\left[\frac{d f(\theta)}{d \theta}\right]  \leq 
  \sum_p \frac{|c_p|^2 \sigma^2}{S_p}. 
  \label{eq:var grad est}
\end{equation}
If we use a total of $S=\sum_p S_p$ shots, then  the minimum variance is obtained by choosing 
\begin{equation}
  S_p = S |c_p| / \norm{\bs c}_1,
  \label{eq:optimal shot allocation}
\end{equation}
where $\bs c$ is the vector with components $c_p$.
With such optimal shot allocation, if we can tolerate 
an error of at most $\varepsilon$, then 
the total number of measurement shots required satisfies 
\begin{equation}
	S \geq \frac{\sigma^2}{\varepsilon^2}\norm{\bs c}_1^2.
 \label{eq:S_optimal}
\end{equation}
Based on the above estimate, we can try to reduce the number of measurement 
shots by working with overparametrized problems and---among the infinitely
many solutions---choose the one with minimum $\norm{\bs c}_1$. 
This problem can then be formally defined as a convex optimization problem,
which can be easily solved using available libraries~\cite{Convex.jl-2014}:
\begin{align}
  \min_{\bs c} \|\bs c\|_1 
  ~~~~~~
  {\rm such~that}
  ~~~~~~
  \sum_{p} c_p e^{\iu \omega \vartheta_p} = \iu \omega,
  ~~~~~~
  \forall \omega \in \Omega,
  \label{eq:convex}
\end{align}
or for symmetric parameter-shift rules
\begin{align}
  \min_{\bs c} \|\bs c\|_1 
  ~~~~~~
  {\rm such~that}
  ~~~~~~
  2 \sum_{p=1}^P c_p \sin(\omega \vartheta_p) = \omega,
  ~~~~~~
  \forall \omega \in \Omega^+.
  \label{eq:convex pos}
\end{align}
Optimization problems like the above have been extensively studied 
in the literature~\cite{chen2001atomic,donoho2008fast}, 
with applications in error correction~\cite{candes2005decoding},
signal reconstruction and magnetic resonance imaging~\cite{candes2006robust},
and compressed sensing~\cite{donoho2006compressed}.
In the quantum setting, they have been 
used to probabilistically interpolate quantum circuits \cite{koczor2024sparse,koczor2024probabilistic}. 

The resulting algorithm is then summarized in Algorithm~\ref{alg:overpsr}.

\begin{algorithm}[H]
  \caption{Overshifted Parameter Shift Rule for a given $\Omega$}
  \label{alg:overpsr}
   \begin{algorithmic}[1]
     \State Fix a ``shift bandwidth'', namely a $B$ such that $|\vartheta|\leq B$. 
     For periodic functions with period $2\pi$ we may set $B=\pi$. 
     \State Fix a suitably large number of shifts. For symmetric shifts we require  that $P \geq |\Omega^+|$, with larger $P$ meaning larger overshifting. 
     \State Define the shifts: e.g.~for symmetric parameter-shift rules and generic $B$, we might set $\vartheta_p=pB/P$. 
     Alternative choices, directly connected to Discrete Fourier Transforms when $B=\pi$, are with
     $\vartheta_p = 2Bp/(2P+1)$ or $\vartheta_p=B(2p-1)/(2P)$.
     \State Check if the linear system \eqref{eq:linear_prob} [or \eqref{eq:linear_prob pos}]
     has at least a solution, e.g.~via Rouch\'e-Capelli's theorem.
     If not, repeat the previous steps with different choices. 
     \State Find the coefficients $\bs c$ by solving either
     Eq.~\eqref{eq:convex} [or \eqref{eq:convex pos}]
     to enforce minimum measurement overhead, or \eqref{eq:convex_smooth}, if we want 
     to enforce continuity. 
     \State Optimize the measurement shots as in Eq.~\eqref{eq:optimal shot allocation} and estimate 
     all values of $f(\theta+\vartheta_p)$ with $S_p$ shots. 
     The final estimate is then given by the weighted average Eq.~\eqref{eq:shift pos}.
   \end{algorithmic}
\end{algorithm}

A simpler, yet non optimal, alternative consists in replacing the $L_1$ norm with the $L_2$ norm, for which 
the resulting convex optimization problem has a closed form solution as
\begin{equation}
  \bs c_{L_2} = D^\sharp \iu \bs\omega,
  \label{eq:L2}
\end{equation}
where $\bs\omega$ is the vectors with components $\omega\in\Omega$, $D$ is the matrix with 
elements $D_{jp} = e^{i\omega_j\vartheta_p }$ with $j=1,\dots,|\Omega|$, and $D^\sharp$ is 
its Moore-Penrose pseudo-inverse. 

\subsection{Towards smooth solutions} 

One possible issue with the solutions in~\eqref{eq:convex} is that they might be 
highly oscillatory, i.e.~with $|c_p-c_q|$ large even when $|\vartheta_p-\vartheta_q|$
is small. This might be undesirable in experimental settings with a non-negligible 
calibration error, namely where precisely calibrating the shifts might be 
challenging.
In order to get smoother solutions, we might then change~\eqref{eq:convex} with 
\begin{align}
  \min_{\bs c} \sum_p |c_{p+1}-c_p|
  ~~~~~~
  \mathrm{such~that}
  ~~~~~~
  \sum_{p} c_p e^{\iu \omega \theta_p} = \iu \omega,
  ~~~~~~
  \forall \omega \in \Omega,
  \label{eq:convex_smooth}
\end{align}
and amend Algorithm~\ref{alg:overpsr} accordingly.
In this way, solutions with highly different neighbouring shifts are penalized. 
Moreover, from the triangle inequality
$ \sum_p |c_{p+1}-c_p| \leq 2 \|\bs c\|_1 $, and from
Eqs.~\eqref{eq:S_optimal} and~\eqref{eq:iterations_bound},  we note that 
the solutions to the above optimization problem might require at most four times 
the number of measurement shots compared with the solutions of~\eqref{eq:convex}, 
which might still be acceptable. 
The convex problem~\eqref{eq:convex_smooth} was already studied in~\cite{candes2006robust}, 
where it was shown to share many of the desired properties of that in Eq.~\eqref{eq:convex}.

\subsection{Continuous limit: stochastic parameter shift rule}\label{sec:continuous}
We have seen that increasing the number of shifts may be beneficial to 
reduce a bound on the number of measurement shots.
In the limit where the number of shifts tends to infinity, we may replace Eqs.~\eqref{eq:shift}
and~\eqref{eq:linear_prob} with their continuous versions
\begin{align}
  \frac{d f(\theta)}{d \theta} &= \int d\vartheta\, c(\vartheta)
  f(\theta+\vartheta), &
  \int d\vartheta\, c(\vartheta) e^{\iu \omega \vartheta} 
  &= \iu \omega, &
  \forall \omega  &\in \Omega,
  \label{eq:continuous}
\end{align}
with a shift density $c(\vartheta)$.
Alternatively, working with positive shifts only we get 
\begin{align}
  \frac{d f(\theta)}{d \theta} &= \int_0^\infty d\vartheta\, c(\vartheta)
  [f(\theta+\vartheta)-f(\theta-\vartheta)], &
  \int_0^\infty d\vartheta\, c(\vartheta) \sin(\omega \vartheta)
                                             &=  \omega, &
  \forall \omega  &\in \Omega^+,
  \label{eq:continuous_pos}
\end{align}
At first the above formulae may seem to be of limited interest,
as it is impossible to experimentally measure a continuous number of 
circuits. However, the above expression is useful to derive a 
stochastic parameter shift rule, which was originally developed for 
gate parametrizations that contain a drift Hamiltonian~\cite{banchi2021measuring}. 

The main motivation behind the stochastic parameter shift rule is stochastic 
gradient descent, which is routinely used in machine learning applications.
In stochastic gradient descent, the optimizer does not employ the exact gradient
but rather an approximation estimated via a finite number of samples. 
To express~\eqref{eq:continuous} in the form of stochastic gradient 
descent, we define $c_\pm(\vartheta)=\max\{0,\pm c(\vartheta)\}$, 
such that $c(\vartheta)=c_+(\vartheta)-c_-(\vartheta)$. 
Notice that this decomposition is necessary also for Eq.~\eqref{eq:continuous_pos}, 
where $c(-\vartheta)=-c(\vartheta)$ but $c(\vartheta)$ may be negative even for 
$\vartheta>0$. 
Then, 
from Eq.~\eqref{eq:continuous} with $\omega=0$, which is always included 
in $\Omega$ thanks to the definitions~\eqref{eq:omega} and~\eqref{eq:omega_shared}, we see 
that $\int d\vartheta c_+(\vartheta) = \int d\vartheta c_-(\vartheta) = \|c\|_1 /2$.
Therefore, we can define two normalized probability 
distributions $p_\pm(\vartheta)=c_\pm(\vartheta) \frac{2}{\|c\|_1}$ and write 
\begin{equation}
  \frac{d f(\theta)}{d \theta} = 
  \frac{\|c\|_1}2\left(
  \ave_{\vartheta_+\sim p_+, \vartheta_-\sim p_-}\left[
  f(\theta+\vartheta_+) - f(\theta+\vartheta_-)\right]\right).
  \label{eq:stoc_grad}
\end{equation}
The gradient can then be estimated by sampling $\vartheta_\pm$ a certain 
number of times from the distributions $p_\pm$ and then estimating 
the abstract average in Eq.~\eqref{eq:stoc_grad} with the empirical 
average using the finite number of samples. 
How to optimally allocate the samples to minimize the variance is discussed 
in Appendix~\ref{app:shot alloc}. 
Remarkably, convergence can be proven even when each gradient is estimated with
a single sample~\cite{bubeck2015convex,gentini2020noise}. Indeed, let $G$ be an upper bound 
on the gradient estimator, then 
stochastic gradient descent converges to a local optimum of $f(\theta)$ 
with an error that is upper bounded by $R \frac{G}{ \sqrt I}$, where $I$ is the number 
of iterations, and $R$ is a constant that depends on the
function and on the parameter space.
Since at each iteration we need to estimate $f(\theta+\theta_\pm)$,
the number of iterations is proportional to the number of measurement shots $S$.
Similarly to Eq.~\eqref{eq:S_optimal}, in order to be $\eta$-close to
the optimum after $I$ iterations, each using 2 measurement shots, we get
\begin{equation}
  S \gtrsim \frac{R^2 G^2}{\eta^2} \propto \|c\|_1^2, 
  \label{eq:iterations_bound}
\end{equation}
where upperbounds $G \propto \|c\|_1$ exist due to Eq.~\eqref{eq:stoc_grad}.
Therefore, we recover the analysis of the previous section: in order 
to minimize the overall number of measurement shots we need 
parameter shift rules that minimize $\|c\|_1$, while also solving 
Eq.~\eqref{eq:continuous}. 

The above steps are summarized in Algorithm~\ref{alg:ops}. Note that the most computationally 
demanding parts, steps 1 and 2, must be done only once. 

\begin{algorithm}[H]
  \caption{Overshifted (Smooth) Stochastic Parameter Shift Rule}
  \label{alg:ops}
   \begin{algorithmic}[1]
     \State Fix the shifts $\vartheta_p$ and find the coefficients $\bs c$, by repeating 
     the steps 1–5 of Algorithm~\ref{alg:overpsr}. 
     \State Define the probability distributions 
     $p_\pm(t) = \max\{0,\pm c_t\} / (\sum_s \max\{0,\pm c_s\})$.
     \State Sample $t_\pm$ from $p_\pm(t)$, e.g.~using 
     Algorithm~\ref{alg:sample_t} from Appendix~\ref{app:derivation}.
     \State Estimate $f(\theta+\vartheta_{t_\pm})$ in a quantum device and 
     call the unbiased estimated result $g_\pm$. 
     \State Define an unbiased estimate of the gradient as $G=(g_+-g_-) \|\bs c\|_1/2$.
     \State Repeat steps 3-5 $S/2$ times and return an average of the estimated $G$. 
   \end{algorithmic}
\end{algorithm}

As we show in Appendix~\ref{app:shot alloc}, the optimal way to estimate each $f(\theta+\vartheta)$ 
in a quantum devices is via a single shot. Namely, given a certain number of total shots $S$, 
the optimal shot allocation is to sample $S$ different values of $\vartheta$ and then 
use a single-shot estimation of each $f(\theta+\vartheta)$. However, for nowadays quantum 
computers, this might be expensive. 
Indeed, for running $f(\theta)$ in a quantum device the abstract circuit must be first compiled 
into native gates and control pulses. If we need to run $f(\theta)$ for many values of $\theta$ 
this complex procedure must be performed each time. 
A simple solution to avoid this problem is to first sample $S$ values of $\vartheta$, and count how many 
times we have sampled the same shift $\vartheta$. If each unique value of $\vartheta$ is found 
$S(\vartheta)$ times, then we can reproduce the same statistics of Algorithm~\ref{alg:ops} 
by estimating $f(\theta+\vartheta)$ with $S(\vartheta)$ shots and then performing a weighted average.  
The resulting procedure is formally described in Algorithm~\ref{alg:avoid recompilation} from 
Appendix~\ref{app:shot alloc}.

\subsection{Uncertainty Principle in Parameter Shifts}\label{sec:uncertainty}
For simplicity, suppose that all the frequencies are commensurate, namely that they can be 
expressed as $\omega_j = \alpha n_j$ for a fixed $\alpha\in\mathbb{R}$ and 
integers $n_j$. If we reabsorb the global $\alpha$ into 
the definition of $\theta$, then $f(\theta)$ is periodic 
with period $2\pi$ and we can focus on functions $c(\vartheta)$ in
Eq.~\eqref{eq:continuous} which share the same periodicity. 
As such, we may expand $c(\vartheta)$ as a discrete Fourier series 
\begin{equation}
  c(\vartheta) = \sum_{n=-\infty}^{\infty} f_n e^{\iu n \vartheta},
\end{equation}
where $f_n^*=f_{-n}$ since $c(\vartheta)$ is real.
Plugging this into Eq.~\eqref{eq:continuous} we get 
\begin{align}
  f_n &= 2\pi \iu n,
      &\forall n\in\Omega,
      \label{eq:f_reduced}
\end{align}
while all the other coefficients $f_n$ with $n\notin \Omega$ can be chosen freely.
The choice $f_n=0$ for $n\notin \Omega $ is not optimal. 
Indeed, solutions to the optimization problem~\eqref{eq:convex} 
have been linked~\cite{donoho2006uncertainty}
to a discrete version of the uncertainty principle~\cite{donoho1989uncertainty,candes2006robust}, 
which essentially states that such solutions 
cannot be sparse in both the real and Fourier domain.
In other terms---if $c(\vartheta)$ is large only for a few values of $\vartheta$---%
as we need to minimize $\|c\|_1$, then the number of non-zero Fourier 
coefficients $f_n$ must be large. 

The interplay between sparsity in one domain and spread in the conjugate domain was discovered in \cite{candes2006robust,candes2005decoding}, where it was considered the problem of reconstructing a signal from highly incomplete frequency information, rather than from sampling at the Nyquist rate. In those settings, when only a small random subset of Fourier coefficients is known, it was shown that for signals that are sparse in the time domain, one can exactly recover the full signal by solving an $L_1$-norm minimization problem, which promotes sparsity.

\section{Analytic approximations}\label{sec:analytics}
In this section, we derive some analytic approximations that are valid for 
any finite set of frequencies.
In Sec.~\ref{sec:numerics} we will subsequently study some particular cases,
and derive simpler and optimal shift rules. 
The general principle is to extend the linear system in Eq.~\eqref{eq:continuous} 
with an interpolating function $I_\Omega(\omega)$,
such that $I_\Omega(\omega)=\omega$ for all $\omega\in \Omega$.
From Eq.~\eqref{eq:continuous} we then get 
\begin{equation}
	c(\vartheta) = \int_{-\infty}^{\infty} \frac{d\omega}{2\pi} e^{-\iu \omega \vartheta} \iu I_\Omega(\omega),
  \label{eq:formal_solution}
\end{equation}
which is valid for all interpolating functions \( I_{\Omega}(\omega)\).
Eq.~\eqref{eq:continuous_pos} is then recovered for odd interpolating functions \( I_{\Omega}(-\omega) = -I_{\Omega}(\omega) \) with
\begin{equation}
	c(\vartheta) = \int_{0}^{\infty} \frac{d\omega}{\pi} \sin(\omega \vartheta) I_\Omega(\omega),
  \label{eq:formal_solution_pos}
\end{equation}
whence we see that $c(-\vartheta)=-c(\vartheta)$.

Solutions involving Dirac deltas 
are not optimal, due to the uncertainty principle, as they have unbounded $\vartheta$. 
Another trivial solution $I_\Omega(\omega)=\omega$ is not optimal as its Fourier transform 
is $\propto \delta'$.
In the following sections we will discuss some approximate solutions 
that can be tuned in order to minimize $\|c\|_1$. 

\subsection{Triangle wave} \label{sec:triangle}
The first interpolating function is based on the triangle wave, which linearly 
interpolates all points in $\Omega$, and also extrapolates over the whole 
infinite domain of $\Omega$ in a zigzag way.
Since it uses the entire infinite set of frequencies,
from the uncertainly principle this solution is expected to be good. 
Moreover, as we will see in Sec.~\ref{sec:numerics}, numerical solutions 
of the problem \eqref{eq:convex} with equispaced frequencies resemble 
a triangular wave, see e.g.~Fig.~\ref{fig:triangular}. 

Triangle waves of period $2T$ and 
amplitudes in $[-1,1]$ have the following functional form
\begin{equation}
  W_{2T}(x) = \frac{8}{\pi^2}\sum_{t=0}^{\infty}\frac{(-1)^t}{(2t+1)^2}
    \sin\left(\frac{(2t+1)\pi x}{T}\right).
\end{equation}
Let $\Lambda \geq |\omega|$ be a number bigger than all beat frequencies in
$\Omega$, e.g.~the bandwidth Eq.~\eqref{eq:bandwidth}.
Then $W_{4\Lambda }(\omega) = \omega /\Lambda$ for all $\omega\in\Omega$ and, accordingly, 
$I_\Omega(\omega) = \Lambda W_{4\Lambda}(\omega)$ defines an interpolating function. 
From Eq.~\eqref{eq:formal_solution} we then get 
\begin{align}
  c(\vartheta)  &= 
  \frac{4 \Lambda }{\pi^2}
  \sum_{t=0}^{\infty}\frac{(-1)^t}{(2t+1)^2}\left[
    \delta\left(\vartheta - \frac{\pi (2t+1)}{2\Lambda}\right) - 
    \delta\left(\vartheta + \frac{\pi (2t+1)}{2\Lambda}\right) 
  \right],
  \label{eq:c_theta}
\end{align}
which results in Algorithm~\ref{alg:triangle_shift}. 
Such an algorithm returns a single unbiased estimate of the gradient. 
Being a stochastic parameter shift rule, we can optimize shot allocations as described 
in Algorithm~\ref{alg:avoid recompilation} from Appendix~\ref{app:shot alloc}. 
The resulting shift rule is given by Algorithm~\ref{alg:triangle_shift_opt}.
\begin{algorithm}[H]
  \caption{Triangle Shift Rule: single-shot unbiased estimator of $f'(\theta)$.}
  \label{alg:triangle_shift}
   \begin{algorithmic}[1]
     \State Fix $\Lambda\geq \max_{\omega\in\Omega}|\omega|$. 
\State Sample $u$ uniformly from $[0,1]\subset\mathbb{R}$.
     \State Repeat the iteration $q_{i} = q_{i-1}+8/\pi^2(2i+1)^{-2}$ with $q_{-1}=0$ while $q_i\leq u$. 
     Let $t$ be the index such that $q_{t-1}\leq u < q_t$. 
     \State Sample a fair coin $p\in\{0,1\}$ 
     and set $\vartheta=(-1)^p \pi (2t+1)/(2\Lambda)$.
     \State Estimate $f(\theta+\vartheta)$ in a quantum device and call the outcome $g$. 
     \State Return $(-1)^{t+p} \Lambda g$. 
   \end{algorithmic}
\end{algorithm}

\begin{algorithm}[H]
  \caption{Cost-Efficient Triangle Shift Rule: unbiased estimator of $f'(\theta)$ with $S$ shots.}
  \label{alg:triangle_shift_opt}
   \begin{algorithmic}[1]
     \State Fix $\Lambda\geq \max_{\omega\in\Omega}|\omega|$. 
     \For{ $s=1,\dots,S$ }
     \State Sample $u$ uniformly from $[0,1]\subset\mathbb{R}$.
     \State Repeat the iteration $q_{i} = q_{i-1}+8/\pi^2(2i+1)^{-2}$ with $q_{-1}=0$ while $q_i\leq u$. 
     \State Let $t$ be the index such that $q_{t-1}\leq u < q_t$. 
     \State Sample a fair coin $p\in\{0,1\}$, set $\vartheta_s=(-1)^p \pi (2t+1)/(2\Lambda)$ and $c_s= \Lambda (-1)^{p+t}$.
     \EndFor
     \State Define $u_i=\theta_{v_i}$ as the set of different values of $\{\theta_s\}_{s=1}^S$, where 
     $v_i$ defines the first occurrence of such shift in the set. Let $n$ be the number of $u_i$, namely 
     the number of distinct $\theta_s$. Let also $S_i = |\{s: u_i = \theta_s\}|$ be the number of occurrences 
     of $u_i$ in the sampled shifts. 
     \State Estimate $f(\theta+u_i)$ in a quantum device using $S_i$ measurement shots and call the outcomes
     $f_{ij}$ where $i=1,\dots,n$ and $j=1,\dots,S_i$. 
     \State Return the average $\sum_{i=1}^n c_{v_i} \sum_{j=1}^{S_i} f_{ij}/S$.
   \end{algorithmic}
\end{algorithm}

From Eq.~\eqref{eq:c_theta} we find $\|c(\vartheta)\|_1 \leq \Lambda$, so the optimal $\Lambda$ is 
indeed the bandwidth Eq.~\eqref{eq:bandwidth}, $\Lambda =\max_{\omega\in\Omega}|\omega|:=\omega_{\textrm{max}}$, 
namely the minimum value compatible with the constraints.
The downside of this method is that $\vartheta$ has infinite support: although values 
with large $t$ have a low probability $\mathcal O(t^{-2})$ to occur, the distribution has 
long tails.
In order to force a more bounded $\vartheta$, it is possible to chose a larger $\Lambda$,
at the expense though of increasing the number of measurements
due to \eqref{eq:S_optimal} and~\eqref{eq:iterations_bound}. 

\subsection{Single zig-zag}\label{sec:linear ramp}
A similar---yet more continuous---solution can be obtained using only the first 
period of the triangle wave, namely with $I_\Omega(\omega) = 
 | t+\Lambda | - | t-\Lambda | -\frac12(| t+2 \Lambda | - | t-2 \Lambda | )$ for which we 
get 
\begin{equation}
  c(\vartheta) = 
  \frac{4 \sin^2 \left(\frac{\theta \Lambda }{2}\right) \sin (\theta \Lambda )}{\pi \theta^2},
  \label{eq:zigzag}
\end{equation}
and $\|c(\vartheta)\|_1\leq 2\Lambda$, so the optimal choice is again $\Lambda=\omega_{\mathrm{max}}$. 
Due to the factor of $2$, this solution has at most twice the $L_1$ norm of the triangle wave. 
However, the above solution is smooth while the triangle wave requires very specific values 
of $\vartheta$. Therefore, this solution might be less affected by imperfect applications 
of the shifts \eqref{eq:continuous} in real quantum hardware. 

The above solution can be expressed as a stochastic parameter shift rule with probability 
\begin{equation}
  p(\vartheta) = \frac{2 \sin^2 \left(\frac{\vartheta \Lambda}{2}\right)}{\pi \Lambda \vartheta^2},
  \label{eq:p_zigzag}
\end{equation}
and cumulative distribution 
\begin{equation}
  F(\vartheta) = \int_{-\infty}^\vartheta d\phi\, p(\phi) = 
  \frac{\pi \Lambda \vartheta + 2 \cos (\Lambda \vartheta) + 2 \Lambda \vartheta \textrm{Si}(\Lambda \vartheta) - 2}{2\pi \Lambda \vartheta },
  \label{eq:F_zigzag}
\end{equation}
where $\textrm{Si}(x) = \int_0^x \sin(t)/t\, dt$.
\begin{algorithm}[H]
  \caption{Zigzag shift rule: unbiased estimator of $f'(\theta)$.}
  \label{alg:zigzag_shift}
   \begin{algorithmic}[1]
     \State Fix $\Lambda \geq \max_{\omega\in\Omega}|\omega|$. 
     \State Use Algorithm~\ref{alg:inverse_sampling} from Appendix~\ref{app:derivation} 
     to sample $\vartheta$ from \eqref{eq:p_zigzag} via \eqref{eq:F_zigzag}.
     \State Estimate $f(\theta+\vartheta)$ in a quantum device and call the outcome $g$. 
     \State Return $ 2\Lambda \sin(\theta\Lambda) g$. 
   \end{algorithmic}
\end{algorithm}
We can then use inverse sampling to sample $\vartheta$ from $p(\vartheta)$,
the resulting algorithm is summarized in Algorithm~\ref{alg:zigzag_shift}.
This algorithm can also be expressed in the language of Sec.~\ref{sec:continuous},
but sampling from the probabilities $p_\pm(\vartheta)$ 
is more complicated due to the lack of an explicit expression like 
\eqref{eq:F_zigzag} for their cumulative distributions. 

The zigzag shift rule has a smooth $c(\vartheta)$, so results may be less affected by 
experimental fluctuations of the parameters. 
However, sampling from the distribution \eqref{eq:p_zigzag} requires finding
the zeros of a non-linear equation, which may be slow.
In order to find parameter shift rules with simpler, classical sampling,
we turn to kernel interpolation in the next section.

\subsection{Kernel Interpolation}\label{sec:kernel}
Gaussian process regression is a popular technique with wide applications in machine learning~\cite{murphy2012machine}.
In the noiseless case, the interpolating function can be expressed as 
\begin{align}
  I_\Omega(\omega) &= \sum_{i}^{|\Omega|} y_{\omega_i} k(\omega-\omega_i) ,
                   &
  y_{\omega_i} &= \sum_{j=1}^{|\Omega|} (K_\Omega^{-1})_{ij} \omega_j,
  \label{eq:kernel_interpol}
\end{align}
where $\omega_i$ are the elements of $\Omega$ and $(K_\Omega)_{ij} = k(\omega_i-\omega_j)$
is a matrix with  $|\Omega|{\times}|\Omega|$ components. 
The positive semidefinite function $k(\omega,\omega')=k(\omega-\omega')$ is called the kernel and,
due to Bochner's theorem, it can be expressed as 
\begin{equation}
  k(\omega-\omega') = \int_{-\infty}^{\infty} d\theta e^{\iu \theta(\omega-\omega')} p(\theta), 
\end{equation}
where $p(\theta)$ is a probability density function. 
Plugging these definitions into Eq.~\eqref{eq:formal_solution} we get 
\begin{equation}
  c(\vartheta) = p(\vartheta) \sum_{j=1}^{|\Omega|} \iu y_{\omega_j} e^{-\iu \omega_j \vartheta} =
  p(\vartheta) \sum_{\omega\in\Omega^+ }2y_\omega \sin(\omega \vartheta),
  \label{eq:kernel_interpol_sol}
\end{equation}
where in the second equation we assume that $y_\omega$ is real with $y_{-\omega} = y_{\omega}$,
which holds for even probability densities  $\rho(\theta)=\rho(-\theta)$, 
namely when the distribution is symmetric around $\theta=0$ and the kernel has real codomain. 
The resulting numerical procedure is described in Algorithm~\ref{alg:kernel}, 
different kernel choices are summarised in Table~\ref{tab:kernels}.

\begin{algorithm}[H]
  \caption{Kernel-based shift rule: unbiased estimator of $f'(\theta)$.}
  \label{alg:kernel}
   \begin{algorithmic}[1]
     \State Choose a suitable distribution $p(\vartheta)$ with the desirable
     properties described in the main text.
     \State Sample $\vartheta$ from $p(\vartheta)$. 
     \State Compute $\lambda = \sum_{\omega\in\Omega^+ }2y_\omega \sin(\omega \vartheta)$.
		 If $\lambda\approx 0$, discard this sample $\vartheta$ and go back to the previous step. 
     \State Estimate $f(\theta+\vartheta)$ in a quantum device and call the outcome $g$. 
     \State Return $ \lambda g$. 
   \end{algorithmic}
\end{algorithm}

\begin{table}[htbp]
  \centering
  {
	\setlength{\tabcolsep}{10pt}
	\renewcommand{\arraystretch}{2.2} 
  \begin{tabular}{ c c c c c }
    \toprule
    Distribution & $p(\vartheta)$ & $k(\omega)$ & $\Delta \vartheta^2$  \\ 
    \midrule
    Normal & $\frac{1}{\sqrt{2\pi\sigma^2}} e^{- \frac{\vartheta^2}{2B^2}}$ 
                         &  $\exp(- B^2\omega^2 / 2)$ 
                         & $B^2$  \\ 
    \hline
    Uniform & $\frac{H(\vartheta+B)-H(\vartheta-B)}{2B}$ 
                         & $\operatorname{sinc}(\omega B/\pi)$ 
                         & $B^2/3$\\ 
    \hline
    Cauchy  & $\frac{\gamma }{\pi}[1 + \left(B \vartheta\right)^2]^{-1}$
                        & $\exp(-|\omega|/B)$  & $\infty$\\ 
    \hline
    Cosine  & $\frac{1}{2B} \left[1+\cos\left(\frac{\vartheta}{B}\,\pi\right)\right]$ 
                        & $\frac{\operatorname{sinc}(B\omega/\pi)}{1 - (B\omega/\pi)^2}$ 
                        &  $B^2\left(\frac{1}{3}-\frac{2}{\pi^2}\right)$\\ 
    \hline
    Wigner  & $\frac2{\pi B^2}\,\sqrt{B^2-\vartheta^2} $ & $ \frac{2}{B\omega} J_1(B \omega) $ 
                        & $\displaystyle \frac{B^2}{4}$\\ 
    \bottomrule
  \end{tabular}}
  \caption{Some distributions with efficient sampling algorithms~\cite{JSSv098i16}, 
    their kernel, and the variance of the parameter shifts.
		All distributions depend on a single tunable hyperparameter $B$. 
    In the table entries, $H$ is the step function, 
    $\operatorname{sinc}(x)=\sin(\pi x)/(\pi x)$ and $J_1(x)$ is a Bessel function. 
  }
  \label{tab:kernels}
\end{table}

Following Eq.~\eqref{eq:kernel_interpol_sol}, we can now choose a probability distribution 
$p(\vartheta)$ that is easy to sample from.
However, the hyperparameters of such distribution must be chosen carefully to make
the coefficients $y_\omega$ small, as $\|c\|_1\leq \|\bs y\|_1$.
Although the sampling part in this algorithm can be made 
very easy with a suitable choice of the distribution $p(\vartheta)$, e.g. a normal or uniform 
distribution, Algorithm~\ref{alg:kernel} has two bottlenecks.
First, depending on $\Omega$, the coefficient
$\lambda = \sum_{\omega\in\Omega^+ }2y_\omega \sin(\omega \vartheta)$
might be close to zero for many values of $\vartheta$.
Although such values can be \textit{classically discarded} without 
calling the quantum hardware, and hence without increasing the measurement cost, they still 
increase the classical computation part. 
Another possibility is use a higher number of measurement shots for the shifts $\vartheta$ 
with larger $\lambda$, as shown in Appendix~\ref{app:shot alloc} -- see in
particular Eq.~\eqref{eq:nonunif shot alloc}. 

The second complication that we observe in numerical experiments is that the kernel 
matrix $K_\Omega$ can be singular. In order to have a guideline about this possibility 
and develop countermeasures, we might use Gershgorin's circle theorem, which basically 
states that the eigenvalues of $K_\Omega$ are within a radius $R_i = \sum_{j\neq i} |k(\omega_i-\omega_j)|$ 
of the diagonal element $k(\omega_i-\omega_i)=1$.
Therefore, to avoid any singularities, it is sufficient (but not necessary) to request that $R_i \ll 1$,
namely that the off-diagonal elements of the kernel matrix are small.
For a minimum separation $\Delta \omega$ among the frequencies in $\Omega$,
this normally implies that the distribution must be broad enough, with   
$\Delta\vartheta\approx (\Delta\omega)^{-1}$. 

\subsection{Constrained solutions}%
\label{sub:Constrained solutions}
Ideally, stochastic parameter shift rule should have the following properties 
\begin{enumerate}[I.]
  \item The shifts should be constrained within a finite interval, $\vartheta \in [-B,B]$. 
  \item It should be simple to sample from their probability distribution. 
  \item The value of $\|c\|_1$ should be as small as possible. 
  \item The rule should be easy to compute even when there are many (exponentially) many 
    frequencies. 
\end{enumerate}
None of the shift rules that we have introduced have all of these properties. The triangle 
shift rule (Algorithm~\ref{alg:triangle_shift}) have basically all of these, with the exception 
of point I. Numerical solutions of Eq.~\eqref{eq:convex} satisfy I-III by design, but 
become challenging when the number of frequencies is very large, namely they don't satisfy IV. 
Kernel methods can have limited support (e.g. the Uniform distribution in Table~\ref{tab:kernels}),
but they require the numerical inversion of a matrix that depends on the number of frequencies, which 
can be large. 

Since we have unveiled the tight connection between shift rules and interpolation, an analytical 
approach to define shift rules that satisfy all of the above constraints might be to use 
band-limited interpolating functions \cite{knab1979interpolation}, which are often based on 
Prolate Spheroidal Wave Functions \cite{thomson2005spectrum,simons2010slepian}, the eigenfunctions of the 
$\operatorname{sinc}$ kernel. As such, these methods are tightly connected to the kernel interpolation with 
$\operatorname{sinc}$ kernels described above---see e.g.~\cite{slepian1978prolate}. 
However, no simple analytical construction exists. 

On the other hand, numerical solutions of Eq.~\eqref{eq:convex pos} show lots of flexibility, as 
we can choose the shifts in the range we want and have guarantees of optimality, within that range.
However, they don't satisfy point IV. Motivated by the triangle shift rule, which only depends on the 
bandwidth Eq.~\eqref{eq:bandwidth}, that can easily be computed in many cases, even when $|\Omega|$ 
is exponentially large, we propose Algorithm~\ref{alg:approx bandwidth} which defines 
an approximate interpolating function in $[-\Lambda,\Lambda]$. 

\begin{algorithm}[H]
  \caption{Approximate shift rule for a given bandwidth} 
  \label{alg:approx bandwidth}
   \begin{algorithmic}[1]
     \State Given the bandwidth $\Lambda$,
     discretize the function $f(\omega)=\omega$ for $\omega \in [-\Lambda,\Lambda]$ and define 
     \begin{align*} 
       \Omega_\Lambda &= \left\{ \frac\ell L \Lambda~~: ~~ \mathrm{for~}\ell=-L,\dots,L, \mathrm{~and~} L\leq P\right\}.
     \end{align*}
     \State Use Algorithm~\ref{alg:overpsr} with the above $\Omega_\Lambda$.
   \end{algorithmic}
\end{algorithm}

\section{Numerical simulations}\label{sec:numerics}
We now test the performance of the different Algorithms proposed in the previous section.

\subsection{Equispaced frequencies}\label{sec:equispaced}
A common scenario, e.g.~when using parameter sharing (Sec.~\ref{sec:sharing})
in qubit-based quantum circuits, or when dealing with photonic quantum circuits 
(Sec.~\ref{sec:photonics}), consists of equispaced frequencies. 
In that setting, 
\begin{equation}
  \Omega = \{-N,-N+1,\dots,-1,0,1,\dots,N-1,N\},
  \label{eq:photonic_frequencies}
\end{equation}
for a given integer $N$.
This case has been extensively considered in the literature~\cite{wierichs2022general,pappalardo2025photonic,hoch2025variational}. 

Since the resulting function is periodic with period $2\pi$ and Eq.~\eqref{eq:linear_prob} can be inverted using 
discrete Fourier transforms, there are basically two main approaches to define the shifts.
Fix any integer $P\geq N$. Then, 
following~\citet{pappalardo2025photonic}, we may set 
\begin{align}
  \vartheta_p &= \frac{2\pi p}{2P+1}, &
  \mathrm{~~for~~} p\in\{-P,-P+1,\dots,P\}.
  \label{eq:odd shifts}
\end{align}
Alternatively, following \citet{wierichs2022general} we may define the shifts as 
\begin{align}
  \vartheta_p &= \frac{\pi (2p-1)}{2P},
  &
  \vartheta_{-p} &= - \vartheta_p, &
  \mathrm{~~for~~} p\in\{1,\dots,P\}.
  \label{eq:even shifts}
\end{align}
In general, there is an odd number of shifts in Eq.~\eqref{eq:odd shifts}, due to the extra ``zero shift''
$\vartheta_0=0$, while the number of shifts in Eq.~\eqref{eq:even shifts} is even, without the ``zero shift''. 
Both choices lead to a linear system with $N$ equations and $P$ variables in Eq.~\eqref{eq:linear_prob pos}, so 
overshifting occurs when $P>N$. 
Note that in this particular case $c_0=0$ for odd shifts, so the extra shift does not play any role 
in the expansion Eq.~\eqref{eq:shift}.

For equispaced shifts the choice of Eq.~\eqref{eq:even shifts} should be preferred,
as it provides analytic shift rules 
when $P=N$, as shown by \citet{wierichs2022general}. 
Suppose though that we were not aware of this explicit solution and that we decided to focus on the suboptimal choice 
Eq.~\eqref{eq:odd shifts}. This is motivated by the fact that, for general
$\Omega$ with possibly incommensurable frequencies, there is no 
explicit guideline to chose a particular set of shifts.

\begin{figure}[t]
  \centering
  \includegraphics[width=0.99\textwidth]{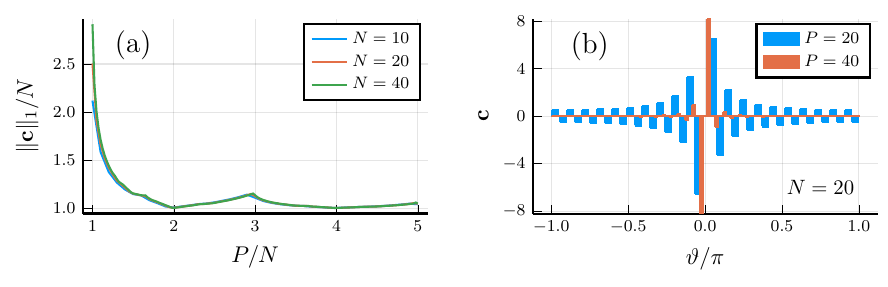}
  \caption{
    (a) Norm of solutions of~\eqref{eq:convex} vs $P/N$ for 
    different values of $N$. 
    (b) Solutions of~\eqref{eq:convex} vs $\vartheta$ for $N=20$ 
    and two values of $P$, $P=20$ (not overshifted) and $P=40$ (overshifted). 
  }
  \label{fig:cnormvsP}
\end{figure}

We use the shifts from Eq.~\eqref{eq:odd shifts} and solve Eq.~\eqref{eq:convex pos} for different values 
of $N$ and $P$. The results are shown in Fig.~\ref{fig:cnormvsP}.
As we see in Fig.~\ref{fig:cnormvsP}(a), 
overshifting reduces $\|\bs c\|_1$, and hence the number of 
measurement shots thanks to Eq.~\eqref{eq:S_optimal}. 
Our results show that, with the subotimal choice of the shifts from Eq.~\eqref{eq:odd shifts},
overshifting is always beneficial to this formulation,
with the optimal at around $P=2N$, where $\|\bs c\|_1 \simeq N$.
Moreover, the relative advantage between the overshifted result for $P=2N$,
and the standard result for $P=2N$ grows for larger $N$,
mostly because the standard solution grows more than linearly.
For instance, according to Fig.~\ref{fig:cnormvsP}(a),
for $N=40$, the number of measurement 
shots required when $P=2N$ is basically one third of that for $P=N$. 
The reason why overshifting is advantageous is apparent from the 
result from Fig.~\ref{fig:cnormvsP}(b).
Indeed, although overshifted rules have more shifts, their coefficients are
much more sparse, with only a few shifts $\theta_p$ having a non-zero value.
On the other hand, when $P=N$ all shifts have a relatively large coefficient $c_p$.

\begin{figure}[t]
  \centering
  \includegraphics[width=0.99\textwidth]{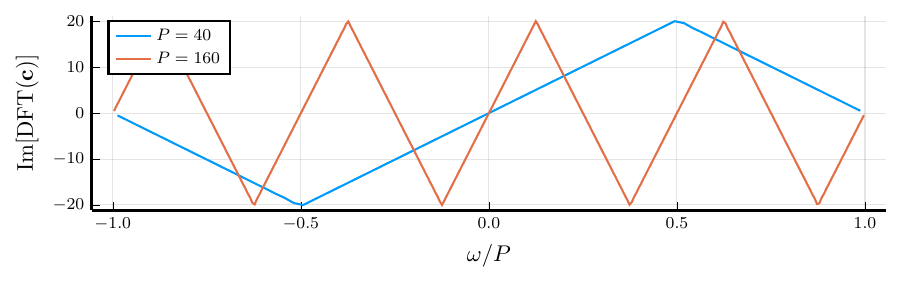}
  \caption{Discrete Fourier Transform of the solutions 
    of \eqref{eq:convex} for $N=20$ and $P=40,160$. The real 
    parts are always zero.
  } 
  \label{fig:triangular}
\end{figure}

In Sec.~\ref{sec:analytics} we discussed several parameter shift rules with an analytic form. 
To have a better understanding about whether the optimal coefficients 
according to problem \eqref{eq:convex}---shown in Fig.~\ref{fig:cnormvsP}(b)---%
can be linked to any of those analytic strategies,
in Fig.~\ref{fig:triangular} we plot the Discrete Fourier Transform (DFT) of such coefficients. 
In that plot, we clearly see a resemblancethatwith triangle waves whose period 
depends on $P$. 
We recall that in Sec.~\ref{sec:uncertainty} we have shown that the Fourier coefficients must be non-zero even for $n\notin\Omega$, namely for $|n|>N$.
From Fig.~\ref{fig:triangular} we see that the solution of \eqref{eq:convex}
basically extends \eqref{eq:f_reduced} over 
$|n|>N$ with the zigzag trend of triangular waves. 
Therefore, we can use the analytic solution of Sec.~\eqref{sec:triangle} 
for any for any $\Lambda \geq N$, with optimal choice $\Lambda=N$. 
Indeed, for this choice $\|c\|_1 \leq N$ and the resulting 
triangle shift rule (Algorithm~\ref{alg:triangle_shift}) 
has the same performance of numerical solutions of problem~\eqref{eq:convex} 
shown in Figure~\ref{fig:cnormvsP}(a). 

The triangle shift rule can be further simplified in this setting.
Indeed, exploiting the periodicity of $f(\theta)$ in Eq.~\eqref{eq:continuous},
in appendix~\ref{app:derivation} we then find that the coefficients \eqref{eq:c_theta} 
can be manipulated to get 
\begin{equation}
  \frac{df(\theta)}{d\theta} = \sum_{t=0}^{N-1}\frac{(-1)^t}{2N\left(1-\cos\vartheta_t\right)}
  \left[f(\theta+\vartheta_t)-f(\theta-\vartheta_t)\right],
  \label{eq:df_conti}
\end{equation}
where $\vartheta_t = \frac{\pi(2t+1)}{2N}$.
The final expression is then equivalent to the one already obtained in~\cite{wierichs2022general,hoch2025variational}.

In summary, we started from the suboptimal choice Eq.~\eqref{eq:odd shifts}, we used the analytic solution 
Eq.~\eqref{eq:c_theta} based on the continuous limit $P\to\infty$, and then by manipulating the resulting 
expression we get Eq.~\eqref{eq:df_conti} which uses the shifts from Eq.~\eqref{eq:even shifts},
which are different from our starting point Eq.~\eqref{eq:odd shifts}.
From our derivation, it is now clear that the choice from~\cite{wierichs2022general,hoch2025variational} 
is optimal to minimize the number of measurement shots. 
It also shows how to use the continuous limit $P\to\infty$ to find the best set of shifts.

Moreover, from Eq.~\eqref{eq:df_conti} we can define a stochastic parameter
shift rule, which is now summarized in Algorithm~\ref{alg:eps}.

\begin{algorithm}[H]
  \caption{Equispaced Stochastic Parameter Shift Rule}
  \label{alg:eps}
   \begin{algorithmic}[1]
     \State Sample $t\in\{0,\dots,N-1\}$ from the probability distribution 
     $p_N(t) = \frac{1}{N^2\left(1-\cos\frac{\pi(2t+1)}{2N}\right)}$,
		 e.g. using Algorithm~\ref{alg:sample_t}.
     \State Compute $\vartheta = \frac{\pi (2t+1)}{2N}$.
     \State Estimate $f(\theta\pm\vartheta)$ in a quantum device and 
     call the unbiased estimated result $g_\pm$. 
     \State Return an unbiased estimate of the gradient $(-1)^t (g_+-g_-) N/2$.
   \end{algorithmic}
\end{algorithm}

\begin{figure}[t]
  \centering
  \includegraphics[width=0.99\textwidth]{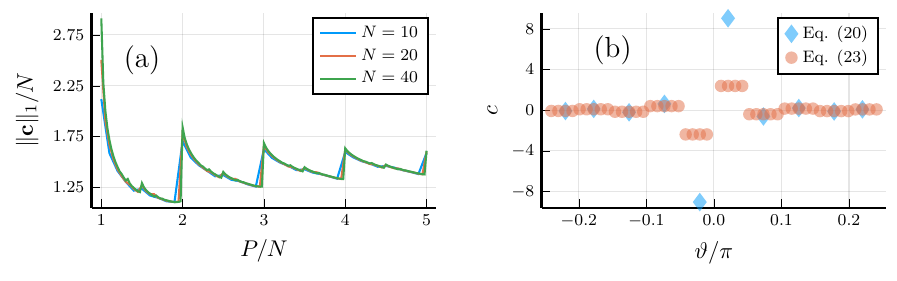}
  \caption{
    (a) Norm of solutions of \eqref{eq:convex_smooth} vs $P/N$ for 
    different values of $N$. 
    (b) Solutions of \eqref{eq:convex} and \eqref{eq:convex_smooth}
    vs $\vartheta$ for $N=20$ and $P=95$. For visual clarity, 
    only the values of $c(\vartheta)$ with $|c| >0.05$ are displayed. 
  }
  \label{fig:cnormvsdiff}
\end{figure}

Finally, in Fig.~\ref{fig:cnormvsdiff} we study the solutions of Eq.~\eqref{eq:convex_smooth}.
In Fig.~\ref{fig:cnormvsdiff}(a), we see that, as expected,
$\|\bs c\|_1$ is typically larger than 
the one obtained from the solution of Eq.~\ref{eq:convex}, shown in Fig.~\ref{fig:cnormvsP}(a). 
In Fig.~\ref{fig:cnormvsdiff}(b) we then show the coefficients 
obtained by solving either Eq.~\eqref{eq:convex} or \eqref{eq:convex_smooth} 
for the same value of $N$ and $P$.
We note that the solutions of of Eq.~\eqref{eq:convex} display only a
non-zero values of $c(\vartheta)$ are a re scattered in $[-\pi,\pi]$.
On the other hand, the solutions of Eq.~\eqref{eq:convex_smooth}
are \textit{clustered}.
Therefore, we expect that these solutions are less affected 
by imperfections in tuning the shifts $\vartheta$.

\subsection{Arbitrary shifts}

While equally-spaced shifts may be the most natural setting to consider, Eq.~\eqref{eq:linear_prob} also allows us to identify other solutions, for example we can obtain a qubit parameter shift rule
\(
    f'(x) \approx - f(x - \frac{\pi}{4}) + 0.707107 f(x) + 0.292893 f(x + \frac{\pi}{2})
\),
although for uniform noise assumptions both the conventional \( \pm \pi/2 \) and the noisier \( \pm \pi/4 \) parameter shift rules outperform it.
Nonetheless this introduces further control that may be valuable to more limited systems or more convoluted noise budgets.

\begin{figure}[t]
	\centering
	\includegraphics{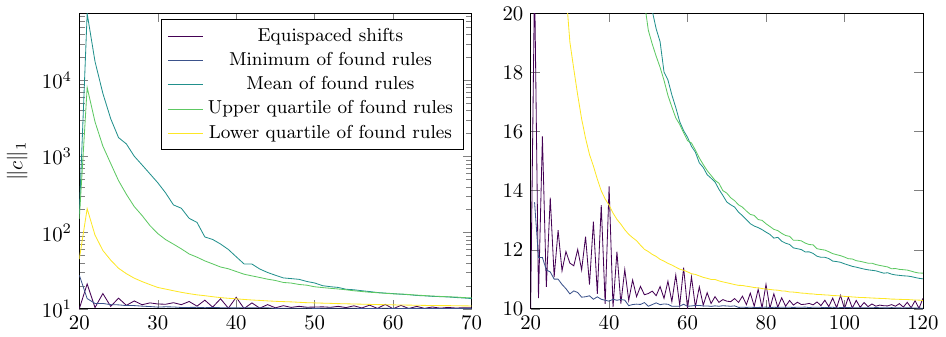}
\caption{Result of 10000 randomly-generated arbitrary shifts for $N=10$ with an equispaced spectrum.
Equispaced shifts are \( \{ (-1+\frac{1}{N_{\mathrm{Shifts}}}) \pi, (-1+\frac{3}{N_{\mathrm{Shifts}}}) \pi, \ldots, (1-\frac{1}{N_{\mathrm{Shifts}}}) \pi \} \), encompassing the \citet{wierichs2022general} and \citet{pappalardo2025photonic} shifts.
}
\label{fig:arb}
\end{figure}

Randomly selecting $n$ shifts from $[-\pi,\pi]^n$, we see in Fig.~\ref{fig:arb} that it is generally feasible to find better rules matching \citet{wierichs2022general}.
Moreover, these randomly selected shifts are more likely to lead to the convex solver finding a valid shift rule, and any such shift rule is less likely to be high cost.

\section{Applications}\label{sec:applic}

\subsection{Photonic Quantum Circuits}\label{sec:photonics}

In photonic quantum circuits parametric gates are normally implemented via 
linear optical elements~\cite{reck1994experimental,clements2016optimal}.
All linear optical components can be expressed as a fixed component, e.g.~a 
50/50 beam splitter, and a phase gate $e^{\iu \theta \hat n}$ where 
$\hat n=\hat a^\dagger \hat a$ is the photon number operator. 
When using single-photon sources, let $N$ be the maximum number of photons
entering into a mode. 
We are then within the settings of Sec.~\ref{sec:equispaced}, where the 
possible frequencies \eqref{eq:photonic_frequencies} are due to the 
allowed eigenvalues of the $\hat n$ operator. 

\subsection{Gaussian States}\label{sec:cv}

\begin{figure}[t]
	\centering
\includegraphics{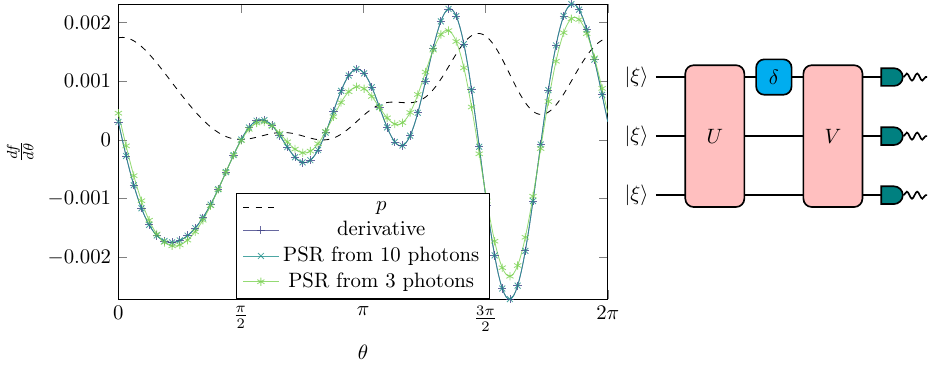}
\caption{Probability of detecting five photons as $(3,1,1)$ in a three-mode interferometer with squeezed vacuum input as a function of a single phase in the first mode.}
\label{fig:gaussian}
\end{figure}

While Gaussian states are in principle infinite-dimensional, with an unbounded spectrum, finite energy states can be approximated by a Fock space expansion, while certain events  can be evaluated exactly from the relevant fixed-number subspaces following Eq.~\eqref{eq:omega_minimal}.

When working with photon number resolving detectors, the proper subspace can be identified from 
the measurement outcomes. For instance, suppose that $\ket\psi$ is a Gaussian state and 
$\ket{\bs n}\bra{\bs n}$ are projectors modelling photon number resolving measurements with outcomes $n_k$ 
on mode $k$. Here $\ket{\bs n} = \ket{n_1,n_2,\dots}$ where $\ket{n_j}$ are Fock states with $n_j$ photons. 
We can reinterpret the probability of getting a particular outcome, $|\bra{\bs n}\psi\rangle|^2$,
as the creation of the multi-mode Fock state $\ket{\bs n}$ followed by the measurement of the Gaussian observable 
$\ket\psi\bra\psi$. In other terms, if measurement results always satisfy $\sum_k n_k \leq N$  for a 
certain cutoff $N$, then the Hilbert space can be approximated as finite dimensional and,
for linear optical circuits, the results of Sec.~\ref{sec:numerics} still apply.

Fig.~\ref{fig:gaussian} shows the probability of detecting a given number of photons at the output of a circuit consisting of a phase shift in one arm of a three-mode Mach--Zehnder-like interferometer, with a squeezed vacuum input.
As the detection is on a Fock state it can---as the observable is equivalent to the outcome of homodyne detection on a fixed photon number state---be solved exactly with a PSR for that number of photons, shift rules for a lower number of photons can still approximate the derivative.

\subsection{Hamiltonian Dynamics of Many-Body Systems } \label{sec:Hamiltonian dynamics}

As an another example application we consider functions obtained by 
letting a many-body quantum system evolve with some Hamiltonian $\hat{H}$, e.g. 
\begin{equation}
  f_H(\theta) = \bra\psi e^{i\hat H \theta} \hat O e^{-i\hat H\theta} \ket\psi,
  \label{eq:Ham evo}
\end{equation}
where $\ket\psi$ is a suitable initial state and $\hat{O}$ is an observable. 
In order to focus on a non-trivial, yet analytically solvable model we focus 
on a spin chain with $L$ qubits interacting via the XY Hamiltonian
\begin{equation}
  \hat H = \frac 14 \sum_{i=1}^{L-1} \left(\hat X_i \hat X_{i+1} + \hat Y_i \hat Y_{i+1}\right), 
\end{equation}
where $\hat{X}_i$ and $\hat{Y}_i$ are Pauli matrices acting on qubit $i$.
The above Hamiltonian can be exactly diagonalized (see e.g.~\cite{banchi2013ballistic})
with energies $E_k = \cos(\pi k/(L+1))$ for $k=1,\dots,L$.
We assume that $\ket\psi$ and $\hat O$ are chosen such that 
$f_H(\theta)$ can be expanded as
\begin{equation}
  f_H(\theta) = \sum_{\omega \in \Omega_H^+} \frac{\cos(\omega \theta)}{\omega}, 
  \label{eq:fH}
\end{equation}
where $\Omega_H^+$ is the set of positive frequencies $\omega = E_k-E_\ell$, with $\omega>0$.

\begin{figure}[t]
  \centering
  \includegraphics[width=0.7\textwidth]{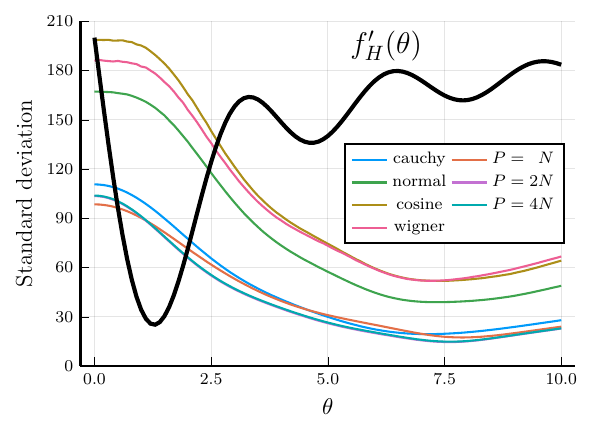}
  \caption{Standard deviation of different estimators of the derivative $f'_{H}(\theta)$, with the $f_H(\theta)$ 
    defined in Eq.~\eqref{eq:fH}. The 
    exact value of $f'_H(\theta)$, rescaled in order to be shown with the same axes of the standard 
    deviation, is shown in black as a reference. The function values at each point, and their 
    variances, are empirically computed using $10^7$ samples. For these numbers, the error 
    in the estimation of $f'_H(\theta)$ is $\approx 200/\sqrt{10^7}$ and all the estimators 
    basically reproduce the true derivative without observable errors in the plot. 
    Cauchy, normal, cosine, and Wigner refer to the analytic estimators discussed
    in Sec.~\ref{sec:analytics} and have been numerically computed using
    Algorithm~\ref{alg:kernel}, while $P=nN$ with $n=1,2,4$ refer to numerical
    solutions of Eq.~\eqref{eq:convex} where $N$ is the number of positive frequencies,
    and the estimators have been numerically computed  using Algorithm~\ref{alg:ops}. 
  }
  \label{fig:cosvariance}
\end{figure}

In Fig.~\ref{fig:cosvariance} we plot the derivative, and the standard deviations of different estimators 
discussed in the previous section for $L=10$ qubits. In such case $N=|\Omega_H^+| = 25$. 
All the estimators were capable of almost perfectly reproducing the value of the derivative, but some 
estimators have larger variances. In particular, those obtained by numerically solving Eq.~\eqref{eq:convex} 
with different values of $P$ show little differences, and the analytical estimator obtained by sampling from
the Cauchy distribution almost matches their performance.
Note though that the latter comes at the price of a non-negligible probability of sampling
large values of $\theta$, since the Cauchy distribution has long tails.
On the other hand, in numerical solutions with $P=N,2N,4N$, the range of
possible $\vartheta$ in Eq.~\eqref{eq:convex} is constrained
by design to a fixed interval, here $[-2\pi,2\pi]$, so their performance can be beaten by other 
methods with a different interval. 

\subsection{Jaynes-Cummings}%
\label{sub:Jaynes-Cummings}

As another example, we focus on the Jaynes-Cummings Hamiltonian, a popular model of
the interactions between an atom and an optical cavity \cite{shore1993jaynes}:
\begin{equation}
\hat H_{\mathrm{JC}} = \frac{\delta}{2}\hat Z + \frac\lambda2 (\hat a^\dagger \hat \sigma_- + \hat a \hat \sigma_+),
\label{eq:jaynes-cummings}
\end{equation}
where $\hat\sigma_\pm = (\hat X\pm \iu \hat Y)/2$, $\hat X,\hat Y, \hat Z$ are the Pauli matrices,
while $\hat a^\dagger$ and $\hat a$ are, respectively, bosonic creation and annihilation operators. 
Since $[\hat Z + \hat a^\dagger \hat a,\hat H]=0$, the Hamiltonian can be diagonalized in each subsector
where $\hat Z + \hat a^\dagger \hat a$ is diagonal and constant. The resulting eigenvalues are 
\begin{align}
  E_n &= \sqrt{\delta^2 + \lambda^2 (n+1)}, & n&=0,1,2,\dots,\infty.
  \label{eq:jc eig}
\end{align}
Although the bandwidth can grow up to infinity, states that are produced in the lab have an energy constraint. 
We can then put an arbitrary cut-off on the photon number, meaning that we can approximate the infinite 
operators $\hat a $ and $\hat a^\dagger$ as $(n+1)$-dimensional matrices.

\begin{figure}[t]
  \centering
  \includegraphics[width=0.99\textwidth]{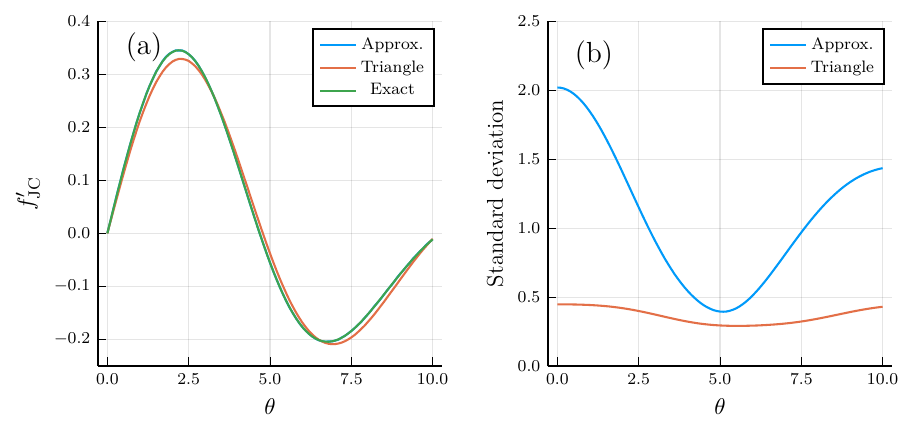}
  \caption{Different estimators of the derivative $f'_{\mathrm{JC}}(\theta)$, 
    with the $f_{\mathrm{JC}}(\theta)$ defined in Eq.~\eqref{eq:f JC} and their standard deviations. 
    We used $\alpha=1$, $\delta=0.2$ and 
    $\lambda=0.5$. 
    We considered two algorithms: the triangle shift rule from Algorithm~\ref{alg:triangle_shift} and
    the Approximate shift rule Algorithm~\ref{alg:approx bandwidth} with $L=100$, $P=1000$ and 
    $\vartheta\in[-\pi,\pi]$.
    Since the model has infinite bandwidth, we define parameter shift rules with a 
    bandwidth Eq.~\eqref{eq:bandwidth} estimated by truncating the model to 10 bosons. Nonetheless, 
    the functions are computed using a larger Hilbert space with a truncation up to 100 bosons, 
    in order to approximate the infinite limit. 
    In (a) the Approximate shift rule is basically indistinguishable from the exact expressions 
    (error at most $10^{-3}$), while the triangle shift rule shows some bias. 
    In (b) we see that the triangle shift rule has a lower standard deviation, as we expect from its optimality,
    though it introduces a bias in (a) due to the wrong estimation of the bandwidth. 
  }
  \label{fig:jcvariance}
\end{figure}

As an example we focus on the function 
\begin{equation}
  f_{\mathrm{JC}}(\theta) = 
  \bra{\psi(\alpha)} e^{\iu \hat H_{\mathrm{JC}}\theta}\hat Z e^{-\iu \hat H_{\mathrm{JC}}\theta}
  \ket{\psi(\alpha)},
  \label{eq:f JC}
\end{equation}
where $\ket{\psi(\alpha)}\propto \left(e^{\alpha\hat a^\dagger}\ket 0\right) \otimes\ket{1}$.
Numerical results are shown in Fig.~\ref{fig:jcvariance}, where we compare the performance 
of the Approximate shift rule and the Triangle shift rule, both of which only require an estimate 
of the bandwidth Eq.~\eqref{eq:bandwidth}. However, since the bandwidth is infinite for this model, we 
perform two approximations. Firstly, we estimate the bandwidth $\Lambda$ by fixing an energy truncation to 10 bosons. 
If we now try to mimic the experimental evaluation of Eq.~\eqref{eq:shift}, even when the coefficients $c_p$ 
are estimated by assuming this energy truncated model, then for each function evaluation 
$f_{\mathrm{JC}}(\theta+\vartheta_p)$ we should basically sum over an infinite number of frequencies. 
Since we cannot perform this limit exactly in numerical simulations, as a second approximation
we estimate each $f_{\mathrm{JC}}(\theta+\vartheta_p)$ with a much larger
cutoff (100 photons) than that used to define the shift rules. 
In this way we try to mimic the experimental setting where we define our derivatives assuming 
an energy constraint, but the functions in Eq.~\eqref{eq:shift} are estimated without such restriction. 

From our numerical solutions, shown in Fig.~\ref{fig:jcvariance}, we see that the both the Approximate 
and Triangle shift rules have a $\mathcal O(1)$ variance, with the triangle shift rule having 
a smaller variance since it has access to an unbounded set of shifts. 
Nonetheless, in Fig.~\ref{fig:jcvariance}(a) we see that the triangle shift rule introduces a bias, 
while the estimate of the approximate shift rule is basically indistinguishable from the exact one. 

The reason behind such bias may be that the triangle shift rule sometimes samples a large shift $\vartheta$,
so possibly the errors due to energy truncation get amplified. 
As a further proof of our intuition, we note that 
 if we use the same truncation to estimate the bandwidth and to evaluate the function 
$f_{\mathrm{JC}}(\theta+\vartheta)$, then both shift rules produce almost exact results, 
with a $\mathcal O(10^{-3})$ error that is compatible with the finite amount of samples ($10^7$).

Our results show that we can play with energy truncation to get a reliable estimate of derivatives 
even for models with incommensurable and infinitely many energies.

\subsection{Parameter sharing}%
\label{sub:Parameter sharing}
Finally, we focus on quantum circuits with dependent parameters. 

\subsubsection{Variational quantum circuits}
As a simple case, consider a problem similar to a Variational Quantum Eigensolver (VQE), 
where the task is to variationally approximate the ground state of a Hamiltonian $\hat H$. 
The variational circuit is constructed as in Eq.~\eqref{eq:psi_theta} with some 
entangling layers $\hat W_\ell$ and fixed rotations. For simplicity, here we assume 
that the rotations are always around the Z axis, namely $\hat H_\ell=\hat Z_{q_\ell}$ 
is a Pauli Z gate on qubit $q_\ell$. In this setting 
\begin{equation}
  f(\bs \theta) = \bra{\psi(\bs\theta)}\hat H\ket{\psi(\bs\theta)}.
\end{equation}
We assume a parametrization 
\begin{equation}
  \theta_i = w_i \theta,
  \label{eq:proj w}
\end{equation}
with some fixed weights $w_i$ 
and the goal is to take gradients with respect to the tunable $\theta$. Approaches like this, namely to project 
the parameter space on a reduced manifold, are 
routinely used in deep neural networks to improve generalization \cite{lotfi2022pac}. 

Gradients with respect to the parametrization Eq.~\eqref{eq:proj w} can be obtained 
using standard parameter shift rule as 
\begin{equation}
\frac{df}{d\theta}= \sum_i \frac{df}{d\theta_i}w_i = \sum_i w_i \left[f(\bs\theta+\frac{\pi}{4} \bs e_i)-f(\bs\theta-\frac{\pi}{4} \bs e_i)\right],
  \label{eq:psr w}
\end{equation}
where $\bs e_i$ is the basis vector with elements $(\bs e_i)_j =\delta_{ij}$.
The above can be rewritten as in Eq.~\eqref{eq:shift} with $\bs c=(w_1,-w_1,w_2,-w_2,\dots)$ 
with $\|\bs c\|_1 = 2\|\bs w\|_1$.

On the other hand, finding the optimal shift rule is prohibitively expensive by working 
with the convex problem \eqref{eq:convex}, as the set of frequencies \eqref{eq:omega_shared} 
increases exponentially with the number of layers. However, since each Z rotation has a frequency $\pm1$, 
as discussed in Sec.~\ref{sec:sharing}, finding the bandwidth is straightforward and we get. 
\begin{equation}
  \Lambda = 2 \|\bs w\|_1.
\end{equation}
Therefore, it is still possible to use Algorithms like the \textit{Triangle shift rule} that 
only depend on the bandwidth and work even for incommensurable frequencies, 
without having to solve complicated equations. From the discussion in Appendix~\ref{app:shot alloc} 
regarding stochastic shift rules and from that in Eq.~\eqref{eq:var grad est} 
about the shot allocation with Eq.~\eqref{eq:psr w}, we find that 
the error coming from the application of the standard parameter shift rule together with the chain rule 
of derivatives is comparable to that of Triangle shift rule.

As discussed previously, one of the disadvantages of the triangle shift rule is that it may sample 
large values of the shift $\vartheta$. Nonetheless, this is not a problem for this example
as Z rotations are periodic, so we can always take $\theta_i = w_i \theta \mathrm{~mod~} 2\pi$, which 
never gets larger than $2\pi$. 

\subsubsection{Structured photonic quantum circuits}\label{sec:sharing:structured}

\begin{figure}[t]
\centering
\includegraphics{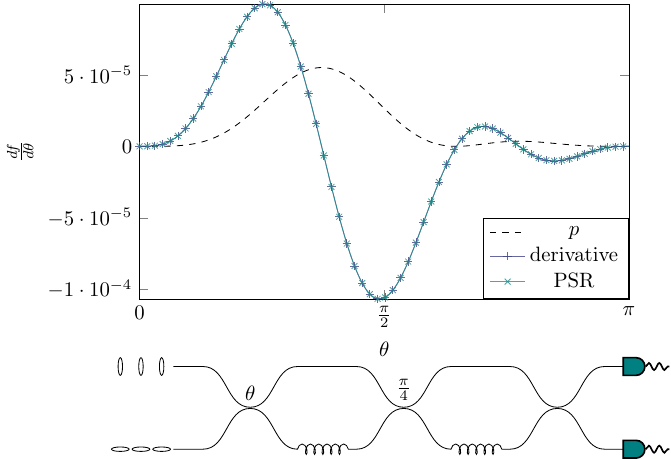}
\caption{Probability of detecting a given set of counts \( \{ n_j \} \) in a 2-spatial mode interferometer, where the first physical beam splitter (that acts on each time-binned pair) varies with transmittance \( \cos\theta \).
The PSR is exact, being evaluated for more energies than \( \sum\limits_j n_j \) }.
\label{fig:shared}
\end{figure}

In order to demonstrate Sec.~\ref{sec:sharing} we consider a small temporarily multiplexed cluster state generation scheme, in the manner of \citet{larsen2019deterministic}, where three temporally spaced pairs pass through the same beam splitter---with a transmittivity tunable through a phase---before a time delay in one mode, a second---fixed and balanced---beam splitter, a further time delay in one mode, and a final---fixed and balanced---beam splitter.
Fig.~\ref{fig:shared} then shows the probability of detecting a specific number of counts in each mode, and the PSR attainable without needing to tune the beam splitter transmittivity within an iteration.

\section{Conclusions}%
\label{sec:Conclusions}

We have generalized the parameter shift rule to circuits 
generated by general gates $e^{\iu \theta \hat H}$ with arbitrary 
Hamiltonians $\hat H$ and arbitrary dimensions of the Hilbert space. 
Our results allow for the estimation of gradients of complex 
quantum circuits and quantum evolutions directly via measurements on the
quantum hardware, with guaranteed minimal overhead. 
We have shown how to define rules that work even when 
the spectrum of the $\hat H$ is unknown or unbounded, e.g.~for infinite dimensional systems. We 
have applied our findings to estimate derivatives of structured quantum 
circuits, Gaussian circuits and circuits made with spin-boson interactions. 

Although not explicitly studied here, 
our results can be readily generalized to other settings. For instance, 
following Ref.~\cite{wierichs2022general} we can extend our rules  
to estimate higher order derivatives by simply changing Eq.~\eqref{eq:convex}. 
Moreover, by trivially adapting the 
stochastic simulation techniques from Ref.~\cite{banchi2021measuring}, 
our rules can be also extended to gates where the parameter 
$\theta$ is one of the parameters in the system Hamiltonian, 
e.g. for gates $e^{\iu (\hat H_0 + \theta \hat V)}$ with arbitrary 
operators $\hat H_0$ and $\hat V$.

Upon completion of this work, a  paper by \citet{lai2025extended} appeared
that proposes to use gradient descent to find new parameter shift rules for
general spectra. Their results complement our work rather than competing with
it. Indeed, their problem is non-convex, meaning that gradient descent is not
guaranteed to find a solution, nor the optimal solution with minimum measurement overhead. On the other hand, our Eq.~\eqref{eq:convex} is
convex and optimal, but requires a fixed choice of the shifts. These two
methods can be combined together. For instance, 
one could use Eq.~\eqref{eq:convex} to get a first estimate
of the shifts, and then fine tune the results (selecting only the shifts with large
enough coefficients) via gradient descent following \cite{lai2025extended}. 
Nonetheless, our
techniques has another advantage, as it can be applied even when the spectrum
of the Hamiltonian that generates the gate is unknown and possibly unbounded, 
as with have shown with the triangle shift rule and with Algorithm~\ref{alg:avoid recompilation}. 

\begin{acknowledgments}
This work is supported by the European Union’s Horizon Europe research and innovation program under the EPIQUE Project (Grant Agreement No.~101135288).
\end{acknowledgments}

\appendix

\section{Shot allocation in stochastic parameter shift rules}\label{app:shot alloc}

We now focus on deciding the optimal shot allocation to estimate derivatives via Eq.~\eqref{eq:stoc_grad}. We first note 
that we can rewrite Eq.~\eqref{eq:stoc_grad} as 
\begin{align}
  \frac{d f(\theta)}{d \theta} &= \|c\|_1 \sum_{s=\pm} p_s \int d\vartheta\, p_s(\vartheta) f(\theta+\vartheta)s
  = \|c\|_1  \ave_{s,\vartheta}[s f(\theta+\vartheta)],
  \label{eq:stoc_grad ex}
\end{align}
where $p_\pm=1/2$. In hardware, $f$ is estimated from the measurement of an observable that here we call $\hat Y$. 
Let us assume that the measured observable has eigenvalues $y$ and eigenvectors $\ket y$. 
From Born's rule, the measurement of $\hat Y$ results in a probability distribution $p(y|\vartheta) = 
|\langle y\ket{\psi(\theta+\vartheta)}|^2$ that depends on $\vartheta$ (and the fixed $\theta$). 
Let $x=(s,\vartheta)$ be a tuple and $e(x)=s \|c\|_1$ be the function that extract the first element of such 
tuple (the sign) and multiplies it by $\|c\|_1$. Then we can rewrite Eq.~\eqref{eq:stoc_grad ex} as 
\begin{align}
  \frac{d f(\theta)}{d \theta} &= \ave_{x\sim p(x)} \ave_{y\sim p(y|x)} [e(x) y],
  \label{eq:stoc_grad xy}
\end{align}
where $p(x) = p_s p_s(\vartheta)$ for $x=(s,\vartheta)$ and $p(y|x)=p(y|\vartheta)$. The problem is then 
reduced to a standard problem in statistics. 

Let $\mu=\mathbb{E}[e(X)Y]$ be the quantity that we want to estimate. Suppose
we draw $n$ distinct $x_i\sim p(x)$. For each $x_i$, we take $m_i\ge 1$
conditionals $y_{ij}\sim p(y|x_i)$ and use the estimator
\begin{equation}
\hat\mu = \frac{1}{n}\sum_{i=1}^n \Bigg(\frac{1}{m_i}\sum_{j=1}^{m_i} e(x_i)y_{ij}\Bigg).
\label{eq:estimator mu}
\end{equation}
By the law of total variance,
\begin{equation}
\mathrm{Var}(\hat\mu)
= \frac{1}{n}\mathrm{Var}\big[e(X)\,Y(X)\big] + \ave \left(
\frac{1}{n^2}\sum_{i=1}^n \frac{e(x_i)^2\,\mathrm{Var}(Y|x_i)}{m_i}\right),
\end{equation}
where $Y(x) = \ave_{y\sim p(y|x)}[y]\equiv f(\theta+\vartheta)$.
The first term denotes the variance due to the uncertainty on $x$, while the second term 
denotes the variance due to the uncertainty coming from quantum measurements, with 
$\mathrm{Var}(Y|x) = \langle \hat Y^2 \rangle_x - \langle \hat Y \rangle_x ^2$.
Clearly the first term does not depend on $m_i$, while the second does. Assuming that 
the total number of measurement shots is constrained by $\sum_{i=1}^n m_i=S$, then we can optimize the 
shot allocation by the following program
\begin{equation}
  \mathrm{minimize}~\left[ \sum_{i=1}^n \frac{w_i^2}{m_i}\right] \quad\text{with}\quad \sum m_i=S,
\qquad\text{where } w_i^2:=e(x_i)^2\,\mathrm{Var}(Y|x_i). 
\end{equation}
Via a Lagrange multiplier we get 
\begin{equation}
m_i \propto w_i = |e(x_i)|\sqrt{\mathrm{Var}(Y|x_i)},
\label{eq:nonunif shot alloc}
\end{equation}
namely we should allocate more samples where  where $|e(x)|$ is large and/or where $Y|x$ is more 
noisy.
For our problem Eq.~\eqref{eq:stoc_grad xy}, $|e(x)|=\|c\|_1$ is constant and the variance of the quantum observable 
can always be bounded by a constant quantity -- e.g., for Pauli observables $Y^2=1$. 
With that suboptimal solution $S=mn$ and 
\begin{equation}
\mathrm{Var}(\hat\mu)
= \frac{m}{S}\mathrm{Var}\big[e(X)\,Y(X)\big] + \frac1{S}
\ave [e(X)^2\,\mathrm{Var}(Y|X)].
\label{eq:varhatmu}
\end{equation}
Since the above function is increasing for larger $m$, the optimal choice is
\begin{align}
  n&=S,  & m=1, 
\end{align}
namely use all the shots to sample $\theta$ and $s$, and use the quantum hardware to estimate 
$f(\theta+\vartheta)$ with a single shot.

As an example, let us consider the case where $\hat Y$ is such that $\hat Y^2=1$. Then 
\begin{align}
  \mathrm{Var}[e(X)Y(X)] &= \|c\|^2_1 \ave_\theta \langle \hat Y\rangle_\theta^2 - \frac{df}{d\theta}^2,  \\
  \ave [e(X)^2\,\mathrm{Var}(Y|X)] &= 
  \|c\|^2_1 \ave_\theta \left( \langle \hat Y^2\rangle_\theta - \langle \hat Y\rangle_\theta^2 \right) = 
  \|c\|^2_1 \left( 1 - \ave_\theta \langle \hat Y\rangle_\theta^2 \right).
\end{align}
Inserting that expression in Eq.~\eqref{eq:varhatmu} we get 
\begin{equation}
  \mathrm{Var}\left[\frac{df}{d\theta}\right]
= \frac{1}{S} \left(\|c\|_1^2 - \frac{df}{d\theta}^2\right),
\end{equation}
which is compatible with Eq.~\eqref{eq:var grad est} with the optimal shot allocation, where $\sigma=1$. 

Finally, we consider how to reshape the above estimator in order to avoid unnecessary circuit 
recompilations, as mentioned at the end of Sec.~\ref{sec:continuous}. 
The resulting algorithm is shown as Algorithm~\ref{alg:avoid recompilation}.
Other estimators might also be more accurate, e.g.~those based on median of means 
or more recent variations \cite{lee2022optimal}, but we found the simple mean to be 
accurate enough in our numerical simulations. 

\begin{algorithm}[H]
  \caption{Estimate Eq.~\eqref{eq:estimator mu} without unnecessary circuit compilations}
  \label{alg:avoid recompilation}
   \begin{algorithmic}[1]
     \State Consider the estimator Eq.~\eqref{eq:estimator mu} with a total
     number of shots $S$ and shot allocation $m_i=1$ and $n=S$.  
     \State Sample $\mathcal X = \{x_i\}_{i=1}^S$ from the joint distribution of parameter shifts and signs $p(x)$. 
     \State Let $ \{\hat x_i\}_{i=1}^n$  be the set different elements of $\mathcal X$ and let $n\leq S$ 
     be its cardinality. 
     \State Let $m_i$ be the number of occurrences of $\hat x_i$ in $\mathcal X$, so $\sum_{i=1}^n m_i=S$. 
     \State Then we can rewrite our estimator as 
     \begin{equation*}
       \hat\mu = \frac{1}{S}\sum_{i=1}^n \sum_{j=1}^{m_i} e(\hat x_i)\hat y_{ij} = 
       \sum_{i=1}^n \frac{m_i}S \frac1{m_i}\sum_{j=1}^{m_i} e(\hat x_i)\hat y_{ij}.
     \end{equation*}
     where now $\hat y_{ij}$ is sampled from $p(y|\hat x_j)$, where 
     $\hat x_j$ are all different. 
     In other terms, we should call the quantum device only with the different values of the sampled 
     shifts $\hat x_j$ and with an adaptive number of shots $m_i$ that depends on the number of occurrences 
     of $\hat x_j$ in $\mathcal X$. Note the differences with Eq.~\eqref{eq:estimator mu}: now there is 
     only a final mean over the total $S$ shots rather than a ``mean of means''. 
     In the second equality we show that the result can also be expressed as a weighted mean of means, 
     with relative weight $m_i/S$. 
   \end{algorithmic}
\end{algorithm}

\section{Some useful algorithms}%
\label{sec:Some useful algorithms}
In this appendix we simply remind a few useful algorithms to sample from complex distributions,
namely Algorithms~\ref{alg:sample_t} and \ref{alg:inverse_sampling}. 

\begin{algorithm}[H]
  \caption{Sample $t\in[0,\dots,T-1]$ from a distribution $p_T(t)$ with $\sum_{t=0}^{T-1}p_T(t)=1$}
  \label{alg:sample_t}
   \begin{algorithmic}[1]
     \State Sample $x$ uniformly from $[0,1]\subset\mathbb R$.
     \State Divide $[0,1]$ into $T$ intervals 
     \begin{equation*}
       [c_0, c_1],[c_1,c_2],\dots,[c_{T-1},c_{T}]
     \end{equation*}
     where $c_0=0$, $c_{T}=1$ and 
     $c_t = \sum_{k=0}^{t-1} p_{T}(t)$ is the comulative distribution. 
     \State Return the index of the interval where $x$ belongs, namely $t$ such 
     that $c_{t}\leq x < c_{t+1}$. 
   \end{algorithmic}
\end{algorithm}

\begin{algorithm}[H]
\caption{Sample from a continuous distribution $p(\theta)$: inverse sampling}
  \label{alg:inverse_sampling}
   \begin{algorithmic}[1]
     \State Sample $u$ uniformly from $[0,1]\subset\mathbb R$.
     \State Find $\theta$ such that $F(\theta) = x$, where $F(\theta)=\int^\theta_{-\infty}p(\vartheta)\,d\vartheta$ 
     is the cumulative distribution. 
     \State Return $\theta$. 
   \end{algorithmic}
\end{algorithm}

\section{Recovering shift rules for equispaced frequencies}
\label{app:derivation}

Starting from Eq.~\eqref{eq:c_theta} we note that, 
after defining $\vartheta_t= \frac{\pi(2t+1)}{2N}$, we have 
$\vartheta_{t+N} = \pi + \vartheta_t$ and 
$\vartheta_{t+2N} = 2\pi - \vartheta_t$.
Therefore, we can write~\eqref{eq:c_theta} as 
\begin{align}
  c(\vartheta)  &= \frac{N}{2}\sum_{t=0}^{2N-1} (-1)^t \eta^{(2N)}_t
	\left[
    \delta\left(\vartheta - \vartheta_t\right)  
		-\delta\left(\vartheta + \vartheta_t\right)
	\right]
\end{align}
where 
\begin{align}
 \eta^{(2N)}_t &= \frac{8}{\pi^2} \sum_{k=0}^\infty \frac{(-1)^{2N k}}{(2t + 4 N k +1)^2}
  = \frac{1}{2\pi^2N^2}\psi^{(1)}\left(\frac{2t+1}{4N}\right),
\end{align}
$\psi^{(1)}(z) = d^2 / dz^2 \log \Gamma(z)$ is the ``trigamma'' function and $\Gamma(z)$ 
is Gamma function, namely the analytic extension to the factorial function. 
Applying this function in the parameter shift rule~\eqref{eq:continuous} we get 
\begin{align}
  \frac{d f(\theta)}{d \theta} &= 
  \frac{N}{2}\sum_{t=0}^{2N-1} (-1)^t \eta^{(2N)}_t \left[
  f(\theta+\vartheta_t)- f(\theta-\vartheta_t)\right]
                             \\&=
  \frac{N}{2}\sum_{t=0}^{N-1} (-1)^t \eta^{(2N)}_t \left[
f(\theta+\vartheta_t)- f(\theta-\vartheta_t)\right] + \\&~~~~~~~~~~~~~\nonumber
  (-1)^{t+N} \eta^{(2N)}_{t+N} \left[
  f(\theta+\vartheta_{t+N})- f(\theta-\vartheta_{t+N})\right]
                             \\&=
  \frac{N}{2}\sum_{t=0}^{N-1} (-1)^t \eta^{(2N)}_t \left[
f(\theta+\vartheta_t)- f(\theta-\vartheta_t)\right] - \\&~~~~~~~~~~~~~\nonumber
  (-1)^{t} \eta^{(2N)}_{2N-1-t} \left[
  f(\theta+\vartheta_{2N-1-t})- f(\theta-\vartheta_{2N-1-t})\right]
                                                     \\ &=
   \frac{N}{2}\sum_{t=0}^{N-1} (-1)^t \left[\eta^{(2N)}_t +\eta^{(2N)}_{2N-1-t}\right]  \left[
f(\theta+\vartheta_t)- f(\theta-\vartheta_t)\right], 
\end{align}
where we used the fact that $f(\theta)=f(\theta+2\pi)$ and $\vartheta_{2N-1-t} = 2\pi-\vartheta_t$. 
Using known functional forms, we then get 
\begin{equation}
  \frac N 2\left[ \eta^{(2N)}_t +\eta^{(2N)}_{2N-1-t}\right] = 
  \frac{1}{4 n \sin ^2\left(\frac{2 \pi  t+\pi }{4 n}\right)}
\end{equation}
and the solution~\eqref{eq:df_conti}.


\begin{thebibliography}{48}%
\makeatletter
\providecommand \@ifxundefined [1]{%
 \@ifx{#1\undefined}
}%
\providecommand \@ifnum [1]{%
 \ifnum #1\expandafter \@firstoftwo
 \else \expandafter \@secondoftwo
 \fi
}%
\providecommand \@ifx [1]{%
 \ifx #1\expandafter \@firstoftwo
 \else \expandafter \@secondoftwo
 \fi
}%
\providecommand \natexlab [1]{#1}%
\providecommand \enquote  [1]{``#1''}%
\providecommand \bibnamefont  [1]{#1}%
\providecommand \bibfnamefont [1]{#1}%
\providecommand \citenamefont [1]{#1}%
\providecommand \href@noop [0]{\@secondoftwo}%
\providecommand \href [0]{\begingroup \@sanitize@url \@href}%
\providecommand \@href[1]{\@@startlink{#1}\@@href}%
\providecommand \@@href[1]{\endgroup#1\@@endlink}%
\providecommand \@sanitize@url [0]{\catcode `\\12\catcode `\$12\catcode
  `\&12\catcode `\#12\catcode `\^12\catcode `\_12\catcode `\%12\relax}%
\providecommand \@@startlink[1]{}%
\providecommand \@@endlink[0]{}%
\providecommand \url  [0]{\begingroup\@sanitize@url \@url }%
\providecommand \@url [1]{\endgroup\@href {#1}{\urlprefix }}%
\providecommand \urlprefix  [0]{URL }%
\providecommand \Eprint [0]{\href }%
\providecommand \doibase [0]{https://doi.org/}%
\providecommand \selectlanguage [0]{\@gobble}%
\providecommand \bibinfo  [0]{\@secondoftwo}%
\providecommand \bibfield  [0]{\@secondoftwo}%
\providecommand \translation [1]{[#1]}%
\providecommand \BibitemOpen [0]{}%
\providecommand \bibitemStop [0]{}%
\providecommand \bibitemNoStop [0]{.\EOS\space}%
\providecommand \EOS [0]{\spacefactor3000\relax}%
\providecommand \BibitemShut  [1]{\csname bibitem#1\endcsname}%
\let\auto@bib@innerbib\@empty
\bibitem [{\citenamefont {Cerezo}\ \emph {et~al.}(2021)\citenamefont {Cerezo},
  \citenamefont {Arrasmith}, \citenamefont {Babbush}, \citenamefont {Benjamin},
  \citenamefont {Endo}, \citenamefont {Fujii}, \citenamefont {McClean},
  \citenamefont {Mitarai}, \citenamefont {Yuan}, \citenamefont {Cincio} \emph
  {et~al.}}]{cerezo2021variational}%
  \BibitemOpen
  \bibfield  {author} {\bibinfo {author} {\bibfnamefont {M.}~\bibnamefont
  {Cerezo}}, \bibinfo {author} {\bibfnamefont {A.}~\bibnamefont {Arrasmith}},
  \bibinfo {author} {\bibfnamefont {R.}~\bibnamefont {Babbush}}, \bibinfo
  {author} {\bibfnamefont {S.~C.}\ \bibnamefont {Benjamin}}, \bibinfo {author}
  {\bibfnamefont {S.}~\bibnamefont {Endo}}, \bibinfo {author} {\bibfnamefont
  {K.}~\bibnamefont {Fujii}}, \bibinfo {author} {\bibfnamefont {J.~R.}\
  \bibnamefont {McClean}}, \bibinfo {author} {\bibfnamefont {K.}~\bibnamefont
  {Mitarai}}, \bibinfo {author} {\bibfnamefont {X.}~\bibnamefont {Yuan}},
  \bibinfo {author} {\bibfnamefont {L.}~\bibnamefont {Cincio}}, \emph
  {et~al.},\ }\bibfield  {title} {\bibinfo {title} {Variational quantum
  algorithms},\ }\href@noop {} {\bibfield  {journal} {\bibinfo  {journal}
  {Nature Reviews Physics}\ }\textbf {\bibinfo {volume} {3}},\ \bibinfo {pages}
  {625} (\bibinfo {year} {2021})}\BibitemShut {NoStop}%
\bibitem [{\citenamefont {Bharti}\ \emph {et~al.}(2022)\citenamefont {Bharti},
  \citenamefont {Cervera-Lierta}, \citenamefont {Kyaw}, \citenamefont {Haug},
  \citenamefont {Alperin-Lea}, \citenamefont {Anand}, \citenamefont {Degroote},
  \citenamefont {Heimonen}, \citenamefont {Kottmann}, \citenamefont {Menke}
  \emph {et~al.}}]{bharti2022noisy}%
  \BibitemOpen
  \bibfield  {author} {\bibinfo {author} {\bibfnamefont {K.}~\bibnamefont
  {Bharti}}, \bibinfo {author} {\bibfnamefont {A.}~\bibnamefont
  {Cervera-Lierta}}, \bibinfo {author} {\bibfnamefont {T.~H.}\ \bibnamefont
  {Kyaw}}, \bibinfo {author} {\bibfnamefont {T.}~\bibnamefont {Haug}}, \bibinfo
  {author} {\bibfnamefont {S.}~\bibnamefont {Alperin-Lea}}, \bibinfo {author}
  {\bibfnamefont {A.}~\bibnamefont {Anand}}, \bibinfo {author} {\bibfnamefont
  {M.}~\bibnamefont {Degroote}}, \bibinfo {author} {\bibfnamefont
  {H.}~\bibnamefont {Heimonen}}, \bibinfo {author} {\bibfnamefont {J.~S.}\
  \bibnamefont {Kottmann}}, \bibinfo {author} {\bibfnamefont {T.}~\bibnamefont
  {Menke}}, \emph {et~al.},\ }\bibfield  {title} {\bibinfo {title} {Noisy
  intermediate-scale quantum algorithms},\ }\href@noop {} {\bibfield  {journal}
  {\bibinfo  {journal} {Reviews of Modern Physics}\ }\textbf {\bibinfo {volume}
  {94}},\ \bibinfo {pages} {015004} (\bibinfo {year} {2022})}\BibitemShut
  {NoStop}%
\bibitem [{\citenamefont {Harrow}\ and\ \citenamefont
  {Napp}(2021)}]{harrow2021low}%
  \BibitemOpen
  \bibfield  {author} {\bibinfo {author} {\bibfnamefont {A.~W.}\ \bibnamefont
  {Harrow}}\ and\ \bibinfo {author} {\bibfnamefont {J.~C.}\ \bibnamefont
  {Napp}},\ }\bibfield  {title} {\bibinfo {title} {Low-depth gradient
  measurements can improve convergence in variational hybrid quantum-classical
  algorithms},\ }\href@noop {} {\bibfield  {journal} {\bibinfo  {journal}
  {Physical Review Letters}\ }\textbf {\bibinfo {volume} {126}},\ \bibinfo
  {pages} {140502} (\bibinfo {year} {2021})}\BibitemShut {NoStop}%
\bibitem [{\citenamefont {Schuld}\ \emph {et~al.}(2019)\citenamefont {Schuld},
  \citenamefont {Bergholm}, \citenamefont {Gogolin}, \citenamefont {Izaac},\
  and\ \citenamefont {Killoran}}]{schuld2019evaluating}%
  \BibitemOpen
  \bibfield  {author} {\bibinfo {author} {\bibfnamefont {M.}~\bibnamefont
  {Schuld}}, \bibinfo {author} {\bibfnamefont {V.}~\bibnamefont {Bergholm}},
  \bibinfo {author} {\bibfnamefont {C.}~\bibnamefont {Gogolin}}, \bibinfo
  {author} {\bibfnamefont {J.}~\bibnamefont {Izaac}},\ and\ \bibinfo {author}
  {\bibfnamefont {N.}~\bibnamefont {Killoran}},\ }\bibfield  {title} {\bibinfo
  {title} {Evaluating analytic gradients on quantum hardware},\ }\href@noop {}
  {\bibfield  {journal} {\bibinfo  {journal} {Physical Review A}\ }\textbf
  {\bibinfo {volume} {99}},\ \bibinfo {pages} {032331} (\bibinfo {year}
  {2019})}\BibitemShut {NoStop}%
\bibitem [{\citenamefont {Wierichs}\ \emph {et~al.}(2022)\citenamefont
  {Wierichs}, \citenamefont {Izaac}, \citenamefont {Wang},\ and\ \citenamefont
  {Lin}}]{wierichs2022general}%
  \BibitemOpen
  \bibfield  {author} {\bibinfo {author} {\bibfnamefont {D.}~\bibnamefont
  {Wierichs}}, \bibinfo {author} {\bibfnamefont {J.}~\bibnamefont {Izaac}},
  \bibinfo {author} {\bibfnamefont {C.}~\bibnamefont {Wang}},\ and\ \bibinfo
  {author} {\bibfnamefont {C.~Y.-Y.}\ \bibnamefont {Lin}},\ }\bibfield  {title}
  {\bibinfo {title} {General parameter-shift rules for quantum gradients},\
  }\href@noop {} {\bibfield  {journal} {\bibinfo  {journal} {Quantum}\ }\textbf
  {\bibinfo {volume} {6}},\ \bibinfo {pages} {677} (\bibinfo {year}
  {2022})}\BibitemShut {NoStop}%
\bibitem [{\citenamefont {Banchi}\ and\ \citenamefont
  {Crooks}(2021)}]{banchi2021measuring}%
  \BibitemOpen
  \bibfield  {author} {\bibinfo {author} {\bibfnamefont {L.}~\bibnamefont
  {Banchi}}\ and\ \bibinfo {author} {\bibfnamefont {G.~E.}\ \bibnamefont
  {Crooks}},\ }\bibfield  {title} {\bibinfo {title} {Measuring analytic
  gradients of general quantum evolution with the stochastic parameter shift
  rule},\ }\href@noop {} {\bibfield  {journal} {\bibinfo  {journal} {Quantum}\
  }\textbf {\bibinfo {volume} {5}},\ \bibinfo {pages} {386} (\bibinfo {year}
  {2021})}\BibitemShut {NoStop}%
\bibitem [{\citenamefont {Wiersema}\ \emph {et~al.}(2024)\citenamefont
  {Wiersema}, \citenamefont {Lewis}, \citenamefont {Wierichs}, \citenamefont
  {Carrasquilla},\ and\ \citenamefont {Killoran}}]{wiersema2024here}%
  \BibitemOpen
  \bibfield  {author} {\bibinfo {author} {\bibfnamefont {R.}~\bibnamefont
  {Wiersema}}, \bibinfo {author} {\bibfnamefont {D.}~\bibnamefont {Lewis}},
  \bibinfo {author} {\bibfnamefont {D.}~\bibnamefont {Wierichs}}, \bibinfo
  {author} {\bibfnamefont {J.}~\bibnamefont {Carrasquilla}},\ and\ \bibinfo
  {author} {\bibfnamefont {N.}~\bibnamefont {Killoran}},\ }\bibfield  {title}
  {\bibinfo {title} {Here comes the su (n): multivariate quantum gates and
  gradients},\ }\href@noop {} {\bibfield  {journal} {\bibinfo  {journal}
  {Quantum}\ }\textbf {\bibinfo {volume} {8}},\ \bibinfo {pages} {1275}
  (\bibinfo {year} {2024})}\BibitemShut {NoStop}%
\bibitem [{\citenamefont {Pappalardo}\ \emph {et~al.}(2025)\citenamefont
  {Pappalardo}, \citenamefont {Emeriau}, \citenamefont {de~Felice},
  \citenamefont {Ventura}, \citenamefont {Jaunin}, \citenamefont {Yeung},
  \citenamefont {Coecke},\ and\ \citenamefont
  {Mansfield}}]{pappalardo2025photonic}%
  \BibitemOpen
  \bibfield  {author} {\bibinfo {author} {\bibfnamefont {A.}~\bibnamefont
  {Pappalardo}}, \bibinfo {author} {\bibfnamefont {P.-E.}\ \bibnamefont
  {Emeriau}}, \bibinfo {author} {\bibfnamefont {G.}~\bibnamefont {de~Felice}},
  \bibinfo {author} {\bibfnamefont {B.}~\bibnamefont {Ventura}}, \bibinfo
  {author} {\bibfnamefont {H.}~\bibnamefont {Jaunin}}, \bibinfo {author}
  {\bibfnamefont {R.}~\bibnamefont {Yeung}}, \bibinfo {author} {\bibfnamefont
  {B.}~\bibnamefont {Coecke}},\ and\ \bibinfo {author} {\bibfnamefont
  {S.}~\bibnamefont {Mansfield}},\ }\bibfield  {title} {\bibinfo {title}
  {Photonic parameter-shift rule: {Enabling} gradient computation for photonic
  quantum computers},\ }\href {https://doi.org/10.1103/PhysRevA.111.032429}
  {\bibfield  {journal} {\bibinfo  {journal} {Physical Review A}\ }\textbf
  {\bibinfo {volume} {111}},\ \bibinfo {pages} {032429} (\bibinfo {year}
  {2025})}\BibitemShut {NoStop}%
\bibitem [{\citenamefont {Hoch}\ \emph {et~al.}(2025)\citenamefont {Hoch},
  \citenamefont {Rodari}, \citenamefont {Giordani}, \citenamefont {Perret},
  \citenamefont {Spagnolo}, \citenamefont {Carvacho}, \citenamefont
  {Pentangelo}, \citenamefont {Piacentini}, \citenamefont {Crespi},
  \citenamefont {Ceccarelli}, \citenamefont {Osellame},\ and\ \citenamefont
  {Sciarrino}}]{hoch2025variational}%
  \BibitemOpen
  \bibfield  {author} {\bibinfo {author} {\bibfnamefont {F.}~\bibnamefont
  {Hoch}}, \bibinfo {author} {\bibfnamefont {G.}~\bibnamefont {Rodari}},
  \bibinfo {author} {\bibfnamefont {T.}~\bibnamefont {Giordani}}, \bibinfo
  {author} {\bibfnamefont {P.}~\bibnamefont {Perret}}, \bibinfo {author}
  {\bibfnamefont {N.}~\bibnamefont {Spagnolo}}, \bibinfo {author}
  {\bibfnamefont {G.}~\bibnamefont {Carvacho}}, \bibinfo {author}
  {\bibfnamefont {C.}~\bibnamefont {Pentangelo}}, \bibinfo {author}
  {\bibfnamefont {S.}~\bibnamefont {Piacentini}}, \bibinfo {author}
  {\bibfnamefont {A.}~\bibnamefont {Crespi}}, \bibinfo {author} {\bibfnamefont
  {F.}~\bibnamefont {Ceccarelli}}, \bibinfo {author} {\bibfnamefont
  {R.}~\bibnamefont {Osellame}},\ and\ \bibinfo {author} {\bibfnamefont
  {F.}~\bibnamefont {Sciarrino}},\ }\bibfield  {title} {\bibinfo {title}
  {Variational approach to photonic quantum circuits via the parameter shift
  rule},\ }\href {https://doi.org/10.1103/PhysRevResearch.7.023227} {\bibfield
  {journal} {\bibinfo  {journal} {Physical Review Research}\ }\textbf {\bibinfo
  {volume} {7}},\ \bibinfo {pages} {023227} (\bibinfo {year}
  {2025})}\BibitemShut {NoStop}%
\bibitem [{\citenamefont {Facelli}\ \emph {et~al.}(2024)\citenamefont
  {Facelli}, \citenamefont {Roberts}, \citenamefont {Wallner}, \citenamefont
  {Makarovskiy}, \citenamefont {Holmes},\ and\ \citenamefont
  {Clements}}]{facelli2024exact}%
  \BibitemOpen
  \bibfield  {author} {\bibinfo {author} {\bibfnamefont {G.}~\bibnamefont
  {Facelli}}, \bibinfo {author} {\bibfnamefont {D.~D.}\ \bibnamefont
  {Roberts}}, \bibinfo {author} {\bibfnamefont {H.}~\bibnamefont {Wallner}},
  \bibinfo {author} {\bibfnamefont {A.}~\bibnamefont {Makarovskiy}}, \bibinfo
  {author} {\bibfnamefont {Z.}~\bibnamefont {Holmes}},\ and\ \bibinfo {author}
  {\bibfnamefont {W.~R.}\ \bibnamefont {Clements}},\ }\bibfield  {title}
  {\bibinfo {title} {Exact gradients for linear optics with single photons},\
  }\href@noop {} {\bibfield  {journal} {\bibinfo  {journal} {arXiv preprint
  arXiv:2409.16369}\ } (\bibinfo {year} {2024})}\BibitemShut {NoStop}%
\bibitem [{\citenamefont {Bauer}\ \emph {et~al.}(2020)\citenamefont {Bauer},
  \citenamefont {Bravyi}, \citenamefont {Motta},\ and\ \citenamefont
  {Chan}}]{bauer2020quantum}%
  \BibitemOpen
  \bibfield  {author} {\bibinfo {author} {\bibfnamefont {B.}~\bibnamefont
  {Bauer}}, \bibinfo {author} {\bibfnamefont {S.}~\bibnamefont {Bravyi}},
  \bibinfo {author} {\bibfnamefont {M.}~\bibnamefont {Motta}},\ and\ \bibinfo
  {author} {\bibfnamefont {G.~K.-L.}\ \bibnamefont {Chan}},\ }\bibfield
  {title} {\bibinfo {title} {Quantum algorithms for quantum chemistry and
  quantum materials science},\ }\href@noop {} {\bibfield  {journal} {\bibinfo
  {journal} {Chemical reviews}\ }\textbf {\bibinfo {volume} {120}},\ \bibinfo
  {pages} {12685} (\bibinfo {year} {2020})}\BibitemShut {NoStop}%
\bibitem [{\citenamefont {Roy}\ \emph {et~al.}(2024)\citenamefont {Roy},
  \citenamefont {Kim}, \citenamefont {Romanenko},\ and\ \citenamefont
  {Grassellino}}]{roy2024qudit}%
  \BibitemOpen
  \bibfield  {author} {\bibinfo {author} {\bibfnamefont {T.}~\bibnamefont
  {Roy}}, \bibinfo {author} {\bibfnamefont {T.}~\bibnamefont {Kim}}, \bibinfo
  {author} {\bibfnamefont {A.}~\bibnamefont {Romanenko}},\ and\ \bibinfo
  {author} {\bibfnamefont {A.}~\bibnamefont {Grassellino}},\ }\href@noop {}
  {\emph {\bibinfo {title} {Qudit-based quantum computing with SRF cavities at
  Fermilab}}},\ \bibinfo {type} {Tech. Rep.}\ (\bibinfo  {institution} {Fermi
  National Accelerator Laboratory (FNAL), Batavia, IL (United States)},\
  \bibinfo {year} {2024})\BibitemShut {NoStop}%
\bibitem [{\citenamefont {Chen}\ \emph {et~al.}(2023)\citenamefont {Chen},
  \citenamefont {Lu}, \citenamefont {Zhang}, \citenamefont {Zhang},
  \citenamefont {Huang}, \citenamefont {Qiao}, \citenamefont {Su},
  \citenamefont {Zhang}, \citenamefont {Zhang}, \citenamefont {Banchi} \emph
  {et~al.}}]{chen2023scalable}%
  \BibitemOpen
  \bibfield  {author} {\bibinfo {author} {\bibfnamefont {W.}~\bibnamefont
  {Chen}}, \bibinfo {author} {\bibfnamefont {Y.}~\bibnamefont {Lu}}, \bibinfo
  {author} {\bibfnamefont {S.}~\bibnamefont {Zhang}}, \bibinfo {author}
  {\bibfnamefont {K.}~\bibnamefont {Zhang}}, \bibinfo {author} {\bibfnamefont
  {G.}~\bibnamefont {Huang}}, \bibinfo {author} {\bibfnamefont
  {M.}~\bibnamefont {Qiao}}, \bibinfo {author} {\bibfnamefont {X.}~\bibnamefont
  {Su}}, \bibinfo {author} {\bibfnamefont {J.}~\bibnamefont {Zhang}}, \bibinfo
  {author} {\bibfnamefont {J.-N.}\ \bibnamefont {Zhang}}, \bibinfo {author}
  {\bibfnamefont {L.}~\bibnamefont {Banchi}}, \emph {et~al.},\ }\bibfield
  {title} {\bibinfo {title} {Scalable and programmable phononic network with
  trapped ions},\ }\href@noop {} {\bibfield  {journal} {\bibinfo  {journal}
  {Nature Physics}\ }\textbf {\bibinfo {volume} {19}},\ \bibinfo {pages} {877}
  (\bibinfo {year} {2023})}\BibitemShut {NoStop}%
\bibitem [{\citenamefont {Crane}\ \emph {et~al.}(2024)\citenamefont {Crane},
  \citenamefont {Smith}, \citenamefont {Tomesh}, \citenamefont {Eickbusch},
  \citenamefont {Martyn}, \citenamefont {K{\"u}hn}, \citenamefont {Funcke},
  \citenamefont {DeMarco}, \citenamefont {Chuang}, \citenamefont {Wiebe} \emph
  {et~al.}}]{crane2024hybrid}%
  \BibitemOpen
  \bibfield  {author} {\bibinfo {author} {\bibfnamefont {E.}~\bibnamefont
  {Crane}}, \bibinfo {author} {\bibfnamefont {K.~C.}\ \bibnamefont {Smith}},
  \bibinfo {author} {\bibfnamefont {T.}~\bibnamefont {Tomesh}}, \bibinfo
  {author} {\bibfnamefont {A.}~\bibnamefont {Eickbusch}}, \bibinfo {author}
  {\bibfnamefont {J.~M.}\ \bibnamefont {Martyn}}, \bibinfo {author}
  {\bibfnamefont {S.}~\bibnamefont {K{\"u}hn}}, \bibinfo {author}
  {\bibfnamefont {L.}~\bibnamefont {Funcke}}, \bibinfo {author} {\bibfnamefont
  {M.~A.}\ \bibnamefont {DeMarco}}, \bibinfo {author} {\bibfnamefont {I.~L.}\
  \bibnamefont {Chuang}}, \bibinfo {author} {\bibfnamefont {N.}~\bibnamefont
  {Wiebe}}, \emph {et~al.},\ }\bibfield  {title} {\bibinfo {title} {Hybrid
  oscillator-qubit quantum processors: Simulating fermions, bosons, and gauge
  fields},\ }\href@noop {} {\bibfield  {journal} {\bibinfo  {journal} {arXiv
  preprint arXiv:2409.03747}\ } (\bibinfo {year} {2024})}\BibitemShut {NoStop}%
\bibitem [{\citenamefont {Larsen}\ \emph {et~al.}(2019)\citenamefont {Larsen},
  \citenamefont {Guo}, \citenamefont {Breum}, \citenamefont
  {Neergaard-Nielsen},\ and\ \citenamefont
  {Andersen}}]{larsen2019deterministic}%
  \BibitemOpen
  \bibfield  {author} {\bibinfo {author} {\bibfnamefont {M.~V.}\ \bibnamefont
  {Larsen}}, \bibinfo {author} {\bibfnamefont {X.}~\bibnamefont {Guo}},
  \bibinfo {author} {\bibfnamefont {C.~R.}\ \bibnamefont {Breum}}, \bibinfo
  {author} {\bibfnamefont {J.~S.}\ \bibnamefont {Neergaard-Nielsen}},\ and\
  \bibinfo {author} {\bibfnamefont {U.~L.}\ \bibnamefont {Andersen}},\
  }\bibfield  {title} {\bibinfo {title} {Deterministic generation of a
  two-dimensional cluster state},\ }\href
  {https://doi.org/10.1126/science.aay4354} {\bibfield  {journal} {\bibinfo
  {journal} {Science}\ }\textbf {\bibinfo {volume} {366}},\ \bibinfo {pages}
  {369} (\bibinfo {year} {2019})}\BibitemShut {NoStop}%
\bibitem [{\citenamefont {Giordani}\ \emph {et~al.}(2023)\citenamefont
  {Giordani}, \citenamefont {Hoch}, \citenamefont {Carvacho}, \citenamefont
  {Spagnolo},\ and\ \citenamefont {Sciarrino}}]{giordani2023integrated}%
  \BibitemOpen
  \bibfield  {author} {\bibinfo {author} {\bibfnamefont {T.}~\bibnamefont
  {Giordani}}, \bibinfo {author} {\bibfnamefont {F.}~\bibnamefont {Hoch}},
  \bibinfo {author} {\bibfnamefont {G.}~\bibnamefont {Carvacho}}, \bibinfo
  {author} {\bibfnamefont {N.}~\bibnamefont {Spagnolo}},\ and\ \bibinfo
  {author} {\bibfnamefont {F.}~\bibnamefont {Sciarrino}},\ }\bibfield  {title}
  {\bibinfo {title} {Integrated photonics in quantum technologies},\
  }\href@noop {} {\bibfield  {journal} {\bibinfo  {journal} {La Rivista del
  Nuovo Cimento}\ }\textbf {\bibinfo {volume} {46}},\ \bibinfo {pages} {71}
  (\bibinfo {year} {2023})}\BibitemShut {NoStop}%
\bibitem [{\citenamefont {Maring}\ \emph {et~al.}(2024)\citenamefont {Maring},
  \citenamefont {Fyrillas}, \citenamefont {Pont}, \citenamefont {Ivanov},
  \citenamefont {Stepanov}, \citenamefont {Margaria}, \citenamefont {Hease},
  \citenamefont {Pishchagin}, \citenamefont {Lema{\^\i}tre}, \citenamefont
  {Sagnes} \emph {et~al.}}]{maring2024versatile}%
  \BibitemOpen
  \bibfield  {author} {\bibinfo {author} {\bibfnamefont {N.}~\bibnamefont
  {Maring}}, \bibinfo {author} {\bibfnamefont {A.}~\bibnamefont {Fyrillas}},
  \bibinfo {author} {\bibfnamefont {M.}~\bibnamefont {Pont}}, \bibinfo {author}
  {\bibfnamefont {E.}~\bibnamefont {Ivanov}}, \bibinfo {author} {\bibfnamefont
  {P.}~\bibnamefont {Stepanov}}, \bibinfo {author} {\bibfnamefont
  {N.}~\bibnamefont {Margaria}}, \bibinfo {author} {\bibfnamefont
  {W.}~\bibnamefont {Hease}}, \bibinfo {author} {\bibfnamefont
  {A.}~\bibnamefont {Pishchagin}}, \bibinfo {author} {\bibfnamefont
  {A.}~\bibnamefont {Lema{\^\i}tre}}, \bibinfo {author} {\bibfnamefont
  {I.}~\bibnamefont {Sagnes}}, \emph {et~al.},\ }\bibfield  {title} {\bibinfo
  {title} {A versatile single-photon-based quantum computing platform},\
  }\href@noop {} {\bibfield  {journal} {\bibinfo  {journal} {Nature Photonics}\
  }\textbf {\bibinfo {volume} {18}},\ \bibinfo {pages} {603} (\bibinfo {year}
  {2024})}\BibitemShut {NoStop}%
\bibitem [{\citenamefont {Aghaee~Rad}\ \emph {et~al.}(2025)\citenamefont
  {Aghaee~Rad}, \citenamefont {Ainsworth}, \citenamefont {Alexander},
  \citenamefont {Altieri}, \citenamefont {Askarani}, \citenamefont {Baby},
  \citenamefont {Banchi}, \citenamefont {Baragiola}, \citenamefont {Bourassa},
  \citenamefont {Chadwick} \emph {et~al.}}]{aghaee2025scaling}%
  \BibitemOpen
  \bibfield  {author} {\bibinfo {author} {\bibfnamefont {H.}~\bibnamefont
  {Aghaee~Rad}}, \bibinfo {author} {\bibfnamefont {T.}~\bibnamefont
  {Ainsworth}}, \bibinfo {author} {\bibfnamefont {R.}~\bibnamefont
  {Alexander}}, \bibinfo {author} {\bibfnamefont {B.}~\bibnamefont {Altieri}},
  \bibinfo {author} {\bibfnamefont {M.}~\bibnamefont {Askarani}}, \bibinfo
  {author} {\bibfnamefont {R.}~\bibnamefont {Baby}}, \bibinfo {author}
  {\bibfnamefont {L.}~\bibnamefont {Banchi}}, \bibinfo {author} {\bibfnamefont
  {B.}~\bibnamefont {Baragiola}}, \bibinfo {author} {\bibfnamefont
  {J.}~\bibnamefont {Bourassa}}, \bibinfo {author} {\bibfnamefont
  {R.}~\bibnamefont {Chadwick}}, \emph {et~al.},\ }\bibfield  {title} {\bibinfo
  {title} {Scaling and networking a modular photonic quantum computer},\
  }\href@noop {} {\bibfield  {journal} {\bibinfo  {journal} {Nature}\ }\textbf
  {\bibinfo {volume} {638}},\ \bibinfo {pages} {912} (\bibinfo {year}
  {2025})}\BibitemShut {NoStop}%
\bibitem [{\citenamefont {Mitarai}\ \emph {et~al.}(2018)\citenamefont
  {Mitarai}, \citenamefont {Negoro}, \citenamefont {Kitagawa},\ and\
  \citenamefont {Fujii}}]{mitarai2018quantum}%
  \BibitemOpen
  \bibfield  {author} {\bibinfo {author} {\bibfnamefont {K.}~\bibnamefont
  {Mitarai}}, \bibinfo {author} {\bibfnamefont {M.}~\bibnamefont {Negoro}},
  \bibinfo {author} {\bibfnamefont {M.}~\bibnamefont {Kitagawa}},\ and\
  \bibinfo {author} {\bibfnamefont {K.}~\bibnamefont {Fujii}},\ }\bibfield
  {title} {\bibinfo {title} {Quantum circuit learning},\ }\href@noop {}
  {\bibfield  {journal} {\bibinfo  {journal} {Physical Review A}\ }\textbf
  {\bibinfo {volume} {98}},\ \bibinfo {pages} {032309} (\bibinfo {year}
  {2018})}\BibitemShut {NoStop}%
\bibitem [{\citenamefont {Knopp}\ \emph {et~al.}(2023)\citenamefont {Knopp},
  \citenamefont {Boberg},\ and\ \citenamefont {Grosser}}]{knopp2023nfft}%
  \BibitemOpen
  \bibfield  {author} {\bibinfo {author} {\bibfnamefont {T.}~\bibnamefont
  {Knopp}}, \bibinfo {author} {\bibfnamefont {M.}~\bibnamefont {Boberg}},\ and\
  \bibinfo {author} {\bibfnamefont {M.}~\bibnamefont {Grosser}},\ }\bibfield
  {title} {\bibinfo {title} {{NFFT.jl}: Generic and fast julia implementation
  of the nonequidistant fast {Fourier} transform},\ }\href@noop {} {\bibfield
  {journal} {\bibinfo  {journal} {SIAM Journal on Scientific Computing}\
  }\textbf {\bibinfo {volume} {45}},\ \bibinfo {pages} {C179} (\bibinfo {year}
  {2023})}\BibitemShut {NoStop}%
\bibitem [{\citenamefont {Ikeda}\ \emph {et~al.}(2024)\citenamefont {Ikeda},
  \citenamefont {Sugiura},\ and\ \citenamefont
  {Polkovnikov}}]{ikeda2024robust}%
  \BibitemOpen
  \bibfield  {author} {\bibinfo {author} {\bibfnamefont {T.~N.}\ \bibnamefont
  {Ikeda}}, \bibinfo {author} {\bibfnamefont {S.}~\bibnamefont {Sugiura}},\
  and\ \bibinfo {author} {\bibfnamefont {A.}~\bibnamefont {Polkovnikov}},\
  }\bibfield  {title} {\bibinfo {title} {Robust effective ground state in a
  nonintegrable floquet quantum circuit},\ }\href@noop {} {\bibfield  {journal}
  {\bibinfo  {journal} {Physical Review Letters}\ }\textbf {\bibinfo {volume}
  {133}},\ \bibinfo {pages} {030401} (\bibinfo {year} {2024})}\BibitemShut
  {NoStop}%
\bibitem [{\citenamefont {Humphreys}\ \emph {et~al.}(2013)\citenamefont
  {Humphreys}, \citenamefont {Metcalf}, \citenamefont {Spring}, \citenamefont
  {Moore}, \citenamefont {Jin}, \citenamefont {Barbieri}, \citenamefont
  {Kolthammer},\ and\ \citenamefont {Walmsley}}]{humphreys2013linear}%
  \BibitemOpen
  \bibfield  {author} {\bibinfo {author} {\bibfnamefont {P.~C.}\ \bibnamefont
  {Humphreys}}, \bibinfo {author} {\bibfnamefont {B.~J.}\ \bibnamefont
  {Metcalf}}, \bibinfo {author} {\bibfnamefont {J.~B.}\ \bibnamefont {Spring}},
  \bibinfo {author} {\bibfnamefont {M.}~\bibnamefont {Moore}}, \bibinfo
  {author} {\bibfnamefont {X.-M.}\ \bibnamefont {Jin}}, \bibinfo {author}
  {\bibfnamefont {M.}~\bibnamefont {Barbieri}}, \bibinfo {author}
  {\bibfnamefont {W.~S.}\ \bibnamefont {Kolthammer}},\ and\ \bibinfo {author}
  {\bibfnamefont {I.~A.}\ \bibnamefont {Walmsley}},\ }\bibfield  {title}
  {\bibinfo {title} {Linear {Optical} {Quantum} {Computing} in a {Single}
  {Spatial} {Mode}},\ }\href {https://doi.org/10.1103/PhysRevLett.111.150501}
  {\bibfield  {journal} {\bibinfo  {journal} {Physical Review Letters}\
  }\textbf {\bibinfo {volume} {111}},\ \bibinfo {pages} {150501} (\bibinfo
  {year} {2013})}\BibitemShut {NoStop}%
\bibitem [{\citenamefont {Madsen}\ \emph {et~al.}(2022)\citenamefont {Madsen},
  \citenamefont {Laudenbach}, \citenamefont {Askarani}, \citenamefont
  {Rortais}, \citenamefont {Vincent}, \citenamefont {Bulmer}, \citenamefont
  {Miatto}, \citenamefont {Neuhaus}, \citenamefont {Helt}, \citenamefont
  {Collins} \emph {et~al.}}]{madsen2022quantum}%
  \BibitemOpen
  \bibfield  {author} {\bibinfo {author} {\bibfnamefont {L.~S.}\ \bibnamefont
  {Madsen}}, \bibinfo {author} {\bibfnamefont {F.}~\bibnamefont {Laudenbach}},
  \bibinfo {author} {\bibfnamefont {M.~F.}\ \bibnamefont {Askarani}}, \bibinfo
  {author} {\bibfnamefont {F.}~\bibnamefont {Rortais}}, \bibinfo {author}
  {\bibfnamefont {T.}~\bibnamefont {Vincent}}, \bibinfo {author} {\bibfnamefont
  {J.~F.}\ \bibnamefont {Bulmer}}, \bibinfo {author} {\bibfnamefont {F.~M.}\
  \bibnamefont {Miatto}}, \bibinfo {author} {\bibfnamefont {L.}~\bibnamefont
  {Neuhaus}}, \bibinfo {author} {\bibfnamefont {L.~G.}\ \bibnamefont {Helt}},
  \bibinfo {author} {\bibfnamefont {M.~J.}\ \bibnamefont {Collins}}, \emph
  {et~al.},\ }\bibfield  {title} {\bibinfo {title} {Quantum computational
  advantage with a programmable photonic processor},\ }\href@noop {} {\bibfield
   {journal} {\bibinfo  {journal} {Nature}\ }\textbf {\bibinfo {volume}
  {606}},\ \bibinfo {pages} {75} (\bibinfo {year} {2022})}\BibitemShut
  {NoStop}%
\bibitem [{\citenamefont {Udell}\ \emph {et~al.}(2014)\citenamefont {Udell},
  \citenamefont {Mohan}, \citenamefont {Zeng}, \citenamefont {Hong},
  \citenamefont {Diamond},\ and\ \citenamefont {Boyd}}]{Convex.jl-2014}%
  \BibitemOpen
  \bibfield  {author} {\bibinfo {author} {\bibfnamefont {M.}~\bibnamefont
  {Udell}}, \bibinfo {author} {\bibfnamefont {K.}~\bibnamefont {Mohan}},
  \bibinfo {author} {\bibfnamefont {D.}~\bibnamefont {Zeng}}, \bibinfo {author}
  {\bibfnamefont {J.}~\bibnamefont {Hong}}, \bibinfo {author} {\bibfnamefont
  {S.}~\bibnamefont {Diamond}},\ and\ \bibinfo {author} {\bibfnamefont
  {S.}~\bibnamefont {Boyd}},\ }\bibfield  {title} {\bibinfo {title} {Convex
  optimization in {J}ulia},\ }in\ \href@noop {} {\emph {\bibinfo {booktitle}
  {Proceedings of the 1st First Workshop for High Performance Technical
  Computing in Dynamic Languages}}}\ (\bibinfo {organization} {IEEE Press},\
  \bibinfo {year} {2014})\ pp.\ \bibinfo {pages} {18--28}\BibitemShut {NoStop}%
\bibitem [{\citenamefont {Chen}\ \emph {et~al.}(2001)\citenamefont {Chen},
  \citenamefont {Donoho},\ and\ \citenamefont {Saunders}}]{chen2001atomic}%
  \BibitemOpen
  \bibfield  {author} {\bibinfo {author} {\bibfnamefont {S.~S.}\ \bibnamefont
  {Chen}}, \bibinfo {author} {\bibfnamefont {D.~L.}\ \bibnamefont {Donoho}},\
  and\ \bibinfo {author} {\bibfnamefont {M.~A.}\ \bibnamefont {Saunders}},\
  }\bibfield  {title} {\bibinfo {title} {Atomic decomposition by basis
  pursuit},\ }\href@noop {} {\bibfield  {journal} {\bibinfo  {journal} {SIAM
  review}\ }\textbf {\bibinfo {volume} {43}},\ \bibinfo {pages} {129} (\bibinfo
  {year} {2001})}\BibitemShut {NoStop}%
\bibitem [{\citenamefont {Donoho}\ and\ \citenamefont
  {Tsaig}(2008)}]{donoho2008fast}%
  \BibitemOpen
  \bibfield  {author} {\bibinfo {author} {\bibfnamefont {D.~L.}\ \bibnamefont
  {Donoho}}\ and\ \bibinfo {author} {\bibfnamefont {Y.}~\bibnamefont {Tsaig}},\
  }\bibfield  {title} {\bibinfo {title} {Fast solution of $l_1$-norm
  minimization problems when the solution may be sparse},\ }\href@noop {}
  {\bibfield  {journal} {\bibinfo  {journal} {IEEE Transactions on Information
  theory}\ }\textbf {\bibinfo {volume} {54}},\ \bibinfo {pages} {4789}
  (\bibinfo {year} {2008})}\BibitemShut {NoStop}%
\bibitem [{\citenamefont {Candes}\ and\ \citenamefont
  {Tao}(2005)}]{candes2005decoding}%
  \BibitemOpen
  \bibfield  {author} {\bibinfo {author} {\bibfnamefont {E.~J.}\ \bibnamefont
  {Candes}}\ and\ \bibinfo {author} {\bibfnamefont {T.}~\bibnamefont {Tao}},\
  }\bibfield  {title} {\bibinfo {title} {Decoding by linear programming},\
  }\href@noop {} {\bibfield  {journal} {\bibinfo  {journal} {IEEE transactions
  on information theory}\ }\textbf {\bibinfo {volume} {51}},\ \bibinfo {pages}
  {4203} (\bibinfo {year} {2005})}\BibitemShut {NoStop}%
\bibitem [{\citenamefont {Cand{\`e}s}\ \emph {et~al.}(2006)\citenamefont
  {Cand{\`e}s}, \citenamefont {Romberg},\ and\ \citenamefont
  {Tao}}]{candes2006robust}%
  \BibitemOpen
  \bibfield  {author} {\bibinfo {author} {\bibfnamefont {E.~J.}\ \bibnamefont
  {Cand{\`e}s}}, \bibinfo {author} {\bibfnamefont {J.}~\bibnamefont
  {Romberg}},\ and\ \bibinfo {author} {\bibfnamefont {T.}~\bibnamefont {Tao}},\
  }\bibfield  {title} {\bibinfo {title} {Robust uncertainty principles: Exact
  signal reconstruction from highly incomplete frequency information},\
  }\href@noop {} {\bibfield  {journal} {\bibinfo  {journal} {IEEE Transactions
  on information theory}\ }\textbf {\bibinfo {volume} {52}},\ \bibinfo {pages}
  {489} (\bibinfo {year} {2006})}\BibitemShut {NoStop}%
\bibitem [{\citenamefont {Donoho}(2006)}]{donoho2006compressed}%
  \BibitemOpen
  \bibfield  {author} {\bibinfo {author} {\bibfnamefont {D.~L.}\ \bibnamefont
  {Donoho}},\ }\bibfield  {title} {\bibinfo {title} {Compressed sensing},\
  }\href@noop {} {\bibfield  {journal} {\bibinfo  {journal} {IEEE Transactions
  on information theory}\ }\textbf {\bibinfo {volume} {52}},\ \bibinfo {pages}
  {1289} (\bibinfo {year} {2006})}\BibitemShut {NoStop}%
\bibitem [{\citenamefont {Koczor}(2024)}]{koczor2024sparse}%
  \BibitemOpen
  \bibfield  {author} {\bibinfo {author} {\bibfnamefont {B.}~\bibnamefont
  {Koczor}},\ }\bibfield  {title} {\bibinfo {title} {Sparse probabilistic
  synthesis of quantum operations},\ }\href
  {https://doi.org/10.1103/PRXQuantum.5.040352} {\bibfield  {journal} {\bibinfo
   {journal} {PRX Quantum}\ }\textbf {\bibinfo {volume} {5}},\ \bibinfo {pages}
  {040352} (\bibinfo {year} {2024})}\BibitemShut {NoStop}%
\bibitem [{\citenamefont {Koczor}\ \emph {et~al.}(2024)\citenamefont {Koczor},
  \citenamefont {Morton},\ and\ \citenamefont
  {Benjamin}}]{koczor2024probabilistic}%
  \BibitemOpen
  \bibfield  {author} {\bibinfo {author} {\bibfnamefont {B.}~\bibnamefont
  {Koczor}}, \bibinfo {author} {\bibfnamefont {J.~J.}\ \bibnamefont {Morton}},\
  and\ \bibinfo {author} {\bibfnamefont {S.~C.}\ \bibnamefont {Benjamin}},\
  }\bibfield  {title} {\bibinfo {title} {Probabilistic interpolation of quantum
  rotation angles},\ }\href@noop {} {\bibfield  {journal} {\bibinfo  {journal}
  {Physical Review Letters}\ }\textbf {\bibinfo {volume} {132}},\ \bibinfo
  {pages} {130602} (\bibinfo {year} {2024})}\BibitemShut {NoStop}%
\bibitem [{\citenamefont {Bubeck}\ \emph {et~al.}(2015)\citenamefont {Bubeck}
  \emph {et~al.}}]{bubeck2015convex}%
  \BibitemOpen
  \bibfield  {author} {\bibinfo {author} {\bibfnamefont {S.}~\bibnamefont
  {Bubeck}} \emph {et~al.},\ }\bibfield  {title} {\bibinfo {title} {Convex
  optimization: Algorithms and complexity},\ }\href@noop {} {\bibfield
  {journal} {\bibinfo  {journal} {Foundations and Trends{\textregistered} in
  Machine Learning}\ }\textbf {\bibinfo {volume} {8}},\ \bibinfo {pages} {231}
  (\bibinfo {year} {2015})}\BibitemShut {NoStop}%
\bibitem [{\citenamefont {Gentini}\ \emph {et~al.}(2020)\citenamefont
  {Gentini}, \citenamefont {Cuccoli}, \citenamefont {Pirandola}, \citenamefont
  {Verrucchi},\ and\ \citenamefont {Banchi}}]{gentini2020noise}%
  \BibitemOpen
  \bibfield  {author} {\bibinfo {author} {\bibfnamefont {L.}~\bibnamefont
  {Gentini}}, \bibinfo {author} {\bibfnamefont {A.}~\bibnamefont {Cuccoli}},
  \bibinfo {author} {\bibfnamefont {S.}~\bibnamefont {Pirandola}}, \bibinfo
  {author} {\bibfnamefont {P.}~\bibnamefont {Verrucchi}},\ and\ \bibinfo
  {author} {\bibfnamefont {L.}~\bibnamefont {Banchi}},\ }\bibfield  {title}
  {\bibinfo {title} {Noise-resilient variational hybrid quantum-classical
  optimization},\ }\href@noop {} {\bibfield  {journal} {\bibinfo  {journal}
  {Physical Review A}\ }\textbf {\bibinfo {volume} {102}},\ \bibinfo {pages}
  {052414} (\bibinfo {year} {2020})}\BibitemShut {NoStop}%
\bibitem [{\citenamefont {Donoho}\ and\ \citenamefont
  {Huo}(2006)}]{donoho2006uncertainty}%
  \BibitemOpen
  \bibfield  {author} {\bibinfo {author} {\bibfnamefont {D.}~\bibnamefont
  {Donoho}}\ and\ \bibinfo {author} {\bibfnamefont {X.}~\bibnamefont {Huo}},\
  }\bibfield  {title} {\bibinfo {title} {Uncertainty principles and ideal
  atomic decomposition},\ }\href@noop {} {\bibfield  {journal} {\bibinfo
  {journal} {IEEE Transactions on Information Theory}\ }\textbf {\bibinfo
  {volume} {47}},\ \bibinfo {pages} {2845} (\bibinfo {year}
  {2006})}\BibitemShut {NoStop}%
\bibitem [{\citenamefont {Donoho}\ and\ \citenamefont
  {Stark}(1989)}]{donoho1989uncertainty}%
  \BibitemOpen
  \bibfield  {author} {\bibinfo {author} {\bibfnamefont {D.~L.}\ \bibnamefont
  {Donoho}}\ and\ \bibinfo {author} {\bibfnamefont {P.~B.}\ \bibnamefont
  {Stark}},\ }\bibfield  {title} {\bibinfo {title} {Uncertainty principles and
  signal recovery},\ }\href@noop {} {\bibfield  {journal} {\bibinfo  {journal}
  {SIAM Journal on Applied Mathematics}\ }\textbf {\bibinfo {volume} {49}},\
  \bibinfo {pages} {906} (\bibinfo {year} {1989})}\BibitemShut {NoStop}%
\bibitem [{\citenamefont {Murphy}(2012)}]{murphy2012machine}%
  \BibitemOpen
  \bibfield  {author} {\bibinfo {author} {\bibfnamefont {K.~P.}\ \bibnamefont
  {Murphy}},\ }\href@noop {} {\emph {\bibinfo {title} {Machine learning: a
  probabilistic perspective}}}\ (\bibinfo  {publisher} {MIT press},\ \bibinfo
  {year} {2012})\BibitemShut {NoStop}%
\bibitem [{\citenamefont {Besançon}\ \emph {et~al.}(2021)\citenamefont
  {Besançon}, \citenamefont {Papamarkou}, \citenamefont {Anthoff},
  \citenamefont {Arslan}, \citenamefont {Byrne}, \citenamefont {Lin},\ and\
  \citenamefont {Pearson}}]{JSSv098i16}%
  \BibitemOpen
  \bibfield  {author} {\bibinfo {author} {\bibfnamefont {M.}~\bibnamefont
  {Besançon}}, \bibinfo {author} {\bibfnamefont {T.}~\bibnamefont
  {Papamarkou}}, \bibinfo {author} {\bibfnamefont {D.}~\bibnamefont {Anthoff}},
  \bibinfo {author} {\bibfnamefont {A.}~\bibnamefont {Arslan}}, \bibinfo
  {author} {\bibfnamefont {S.}~\bibnamefont {Byrne}}, \bibinfo {author}
  {\bibfnamefont {D.}~\bibnamefont {Lin}},\ and\ \bibinfo {author}
  {\bibfnamefont {J.}~\bibnamefont {Pearson}},\ }\bibfield  {title} {\bibinfo
  {title} {Distributions.jl: Definition and modeling of probability
  distributions in the juliastats ecosystem},\ }\href
  {https://doi.org/10.18637/jss.v098.i16} {\bibfield  {journal} {\bibinfo
  {journal} {Journal of Statistical Software}\ }\textbf {\bibinfo {volume}
  {98}},\ \bibinfo {pages} {1} (\bibinfo {year} {2021})}\BibitemShut {NoStop}%
\bibitem [{\citenamefont {Knab}(1979)}]{knab1979interpolation}%
  \BibitemOpen
  \bibfield  {author} {\bibinfo {author} {\bibfnamefont {J.}~\bibnamefont
  {Knab}},\ }\bibfield  {title} {\bibinfo {title} {Interpolation of
  band-limited functions using the approximate prolate series},\ }\href@noop {}
  {\bibfield  {journal} {\bibinfo  {journal} {IEEE transactions on Information
  Theory}\ }\textbf {\bibinfo {volume} {25}},\ \bibinfo {pages} {717} (\bibinfo
  {year} {1979})}\BibitemShut {NoStop}%
\bibitem [{\citenamefont {Thomson}(2005)}]{thomson2005spectrum}%
  \BibitemOpen
  \bibfield  {author} {\bibinfo {author} {\bibfnamefont {D.~J.}\ \bibnamefont
  {Thomson}},\ }\bibfield  {title} {\bibinfo {title} {Spectrum estimation and
  harmonic analysis},\ }\href@noop {} {\bibfield  {journal} {\bibinfo
  {journal} {Proceedings of the IEEE}\ }\textbf {\bibinfo {volume} {70}},\
  \bibinfo {pages} {1055} (\bibinfo {year} {2005})}\BibitemShut {NoStop}%
\bibitem [{\citenamefont {Simons}(2010)}]{simons2010slepian}%
  \BibitemOpen
  \bibfield  {author} {\bibinfo {author} {\bibfnamefont {F.~J.}\ \bibnamefont
  {Simons}},\ }\bibfield  {title} {\bibinfo {title} {Slepian functions and
  their use in signal estimation and spectral analysis},\ }in\ \href@noop {}
  {\emph {\bibinfo {booktitle} {Handbook of geomathematics}}}\ (\bibinfo
  {publisher} {Springer},\ \bibinfo {year} {2010})\ pp.\ \bibinfo {pages}
  {891--923}\BibitemShut {NoStop}%
\bibitem [{\citenamefont {Slepian}(1978)}]{slepian1978prolate}%
  \BibitemOpen
  \bibfield  {author} {\bibinfo {author} {\bibfnamefont {D.}~\bibnamefont
  {Slepian}},\ }\bibfield  {title} {\bibinfo {title} {Prolate spheroidal wave
  functions, fourier analysis, and uncertainty—v: The discrete case},\
  }\href@noop {} {\bibfield  {journal} {\bibinfo  {journal} {Bell System
  Technical Journal}\ }\textbf {\bibinfo {volume} {57}},\ \bibinfo {pages}
  {1371} (\bibinfo {year} {1978})}\BibitemShut {NoStop}%
\bibitem [{\citenamefont {Reck}\ \emph {et~al.}(1994)\citenamefont {Reck},
  \citenamefont {Zeilinger}, \citenamefont {Bernstein},\ and\ \citenamefont
  {Bertani}}]{reck1994experimental}%
  \BibitemOpen
  \bibfield  {author} {\bibinfo {author} {\bibfnamefont {M.}~\bibnamefont
  {Reck}}, \bibinfo {author} {\bibfnamefont {A.}~\bibnamefont {Zeilinger}},
  \bibinfo {author} {\bibfnamefont {H.~J.}\ \bibnamefont {Bernstein}},\ and\
  \bibinfo {author} {\bibfnamefont {P.}~\bibnamefont {Bertani}},\ }\bibfield
  {title} {\bibinfo {title} {Experimental realization of any discrete unitary
  operator},\ }\href@noop {} {\bibfield  {journal} {\bibinfo  {journal}
  {Physical review letters}\ }\textbf {\bibinfo {volume} {73}},\ \bibinfo
  {pages} {58} (\bibinfo {year} {1994})}\BibitemShut {NoStop}%
\bibitem [{\citenamefont {Clements}\ \emph {et~al.}(2016)\citenamefont
  {Clements}, \citenamefont {Humphreys}, \citenamefont {Metcalf}, \citenamefont
  {Kolthammer},\ and\ \citenamefont {Walmsley}}]{clements2016optimal}%
  \BibitemOpen
  \bibfield  {author} {\bibinfo {author} {\bibfnamefont {W.~R.}\ \bibnamefont
  {Clements}}, \bibinfo {author} {\bibfnamefont {P.~C.}\ \bibnamefont
  {Humphreys}}, \bibinfo {author} {\bibfnamefont {B.~J.}\ \bibnamefont
  {Metcalf}}, \bibinfo {author} {\bibfnamefont {W.~S.}\ \bibnamefont
  {Kolthammer}},\ and\ \bibinfo {author} {\bibfnamefont {I.~A.}\ \bibnamefont
  {Walmsley}},\ }\bibfield  {title} {\bibinfo {title} {Optimal design for
  universal multiport interferometers},\ }\href@noop {} {\bibfield  {journal}
  {\bibinfo  {journal} {Optica}\ }\textbf {\bibinfo {volume} {3}},\ \bibinfo
  {pages} {1460} (\bibinfo {year} {2016})}\BibitemShut {NoStop}%
\bibitem [{\citenamefont {Banchi}(2013)}]{banchi2013ballistic}%
  \BibitemOpen
  \bibfield  {author} {\bibinfo {author} {\bibfnamefont {L.}~\bibnamefont
  {Banchi}},\ }\bibfield  {title} {\bibinfo {title} {Ballistic quantum state
  transfer in spin chains: General theory for quasi-free models and arbitrary
  initial states},\ }\href@noop {} {\bibfield  {journal} {\bibinfo  {journal}
  {The European Physical Journal Plus}\ }\textbf {\bibinfo {volume} {128}},\
  \bibinfo {pages} {1} (\bibinfo {year} {2013})}\BibitemShut {NoStop}%
\bibitem [{\citenamefont {Shore}\ and\ \citenamefont
  {Knight}(1993)}]{shore1993jaynes}%
  \BibitemOpen
  \bibfield  {author} {\bibinfo {author} {\bibfnamefont {B.~W.}\ \bibnamefont
  {Shore}}\ and\ \bibinfo {author} {\bibfnamefont {P.~L.}\ \bibnamefont
  {Knight}},\ }\bibfield  {title} {\bibinfo {title} {The jaynes-cummings
  model},\ }\href@noop {} {\bibfield  {journal} {\bibinfo  {journal} {Journal
  of Modern Optics}\ }\textbf {\bibinfo {volume} {40}},\ \bibinfo {pages}
  {1195} (\bibinfo {year} {1993})}\BibitemShut {NoStop}%
\bibitem [{\citenamefont {Lotfi}\ \emph {et~al.}(2022)\citenamefont {Lotfi},
  \citenamefont {Finzi}, \citenamefont {Kapoor}, \citenamefont {Potapczynski},
  \citenamefont {Goldblum},\ and\ \citenamefont {Wilson}}]{lotfi2022pac}%
  \BibitemOpen
  \bibfield  {author} {\bibinfo {author} {\bibfnamefont {S.}~\bibnamefont
  {Lotfi}}, \bibinfo {author} {\bibfnamefont {M.}~\bibnamefont {Finzi}},
  \bibinfo {author} {\bibfnamefont {S.}~\bibnamefont {Kapoor}}, \bibinfo
  {author} {\bibfnamefont {A.}~\bibnamefont {Potapczynski}}, \bibinfo {author}
  {\bibfnamefont {M.}~\bibnamefont {Goldblum}},\ and\ \bibinfo {author}
  {\bibfnamefont {A.~G.}\ \bibnamefont {Wilson}},\ }\bibfield  {title}
  {\bibinfo {title} {Pac-bayes compression bounds so tight that they can
  explain generalization},\ }\href@noop {} {\bibfield  {journal} {\bibinfo
  {journal} {Advances in Neural Information Processing Systems}\ }\textbf
  {\bibinfo {volume} {35}},\ \bibinfo {pages} {31459} (\bibinfo {year}
  {2022})}\BibitemShut {NoStop}%
\bibitem [{\citenamefont {Lai}\ \emph {et~al.}(2025)\citenamefont {Lai},
  \citenamefont {Hu}, \citenamefont {An},\ and\ \citenamefont
  {Wen}}]{lai2025extended}%
  \BibitemOpen
  \bibfield  {author} {\bibinfo {author} {\bibfnamefont {Z.}~\bibnamefont
  {Lai}}, \bibinfo {author} {\bibfnamefont {J.}~\bibnamefont {Hu}}, \bibinfo
  {author} {\bibfnamefont {D.}~\bibnamefont {An}},\ and\ \bibinfo {author}
  {\bibfnamefont {Z.}~\bibnamefont {Wen}},\ }\bibfield  {title} {\bibinfo
  {title} {Extended parameter shift rules with minimal derivative variance for
  parameterized quantum circuits},\ }\href@noop {} {\bibfield  {journal}
  {\bibinfo  {journal} {arXiv preprint arXiv:2508.08802}\ } (\bibinfo {year}
  {2025})}\BibitemShut {NoStop}%
\bibitem [{\citenamefont {Lee}\ and\ \citenamefont
  {Valiant}(2022)}]{lee2022optimal}%
  \BibitemOpen
  \bibfield  {author} {\bibinfo {author} {\bibfnamefont {J.~C.}\ \bibnamefont
  {Lee}}\ and\ \bibinfo {author} {\bibfnamefont {P.}~\bibnamefont {Valiant}},\
  }\bibfield  {title} {\bibinfo {title} {Optimal sub-gaussian mean estimation
  in $\mathbb{R}$},\ }in\ \href@noop {} {\emph {\bibinfo {booktitle} {2021 IEEE
  62nd Annual Symposium on Foundations of Computer Science (FOCS)}}}\ (\bibinfo
  {organization} {IEEE},\ \bibinfo {year} {2022})\ pp.\ \bibinfo {pages}
  {672--683}\BibitemShut {NoStop}%
\end{thebibliography}
\end{document}